\documentclass[nofootinbib,aps,preprint,floats,floatfix,amsmath,amssymb,longbibliography,secnumarabic,superscriptaddress,preprintnumbers]{revtex4-1}
\usepackage{natbib}
\usepackage{tcolorbox}
\usepackage{amssymb}
\usepackage{amsmath}
\usepackage{amsfonts}
\usepackage{graphicx}
\usepackage[hang,tight,raggedright]{subfigure}
\usepackage{multirow}
\usepackage{subfigure}
\usepackage{float}
\usepackage{graphics}
\usepackage{slashed}
\usepackage{color}
\usepackage{bm}
\usepackage{braket}
\usepackage{psfrag}
\usepackage{booktabs}
\usepackage{latexsym}
\usepackage[colorlinks,linkcolor=blue,anchorcolor=blue,citecolor=blue,urlcolor=blue,]{hyperref}
\allowdisplaybreaks

\begin{document}
		
\title{Sound velocity effects on the phase transition gravitational wave spectrum in the sound shell model }

\author{Xiao Wang}	
\email{wangxiao7@mail.sysu.edu.cn}

\author{Fa Peng Huang}
\email{Corresponding Author.  huangfp8@mail.sysu.edu.cn}

\affiliation{MOE Key Laboratory of TianQin Mission, TianQin Research Center for
Gravitational Physics $\&$ School of Physics and Astronomy, Frontiers
Science Center for TianQin, Gravitational Wave Research Center of
CNSA, Sun Yat-sen University (Zhuhai Campus), Zhuhai 519082, China}

\author{Yongping Li}
\email{liyp@ihep.ac.cn}
\affiliation{Theoretical Physics Division, Institute of High Energy Physics, Chinese Academy of Sciences,
	19B Yuquan Road, Shijingshan District, Beijing 100049, China}
\affiliation{School of Physics, University of Chinese Academy of Sciences, Beijing 100049, China}

\begin{abstract}	
A cosmological first-order phase transition gravitational wave could provide a novel approach to studying the early Universe.  
In most cases, the acoustic gravitational wave from the sound wave mechanism is dominant.
Considering different sound velocities in symmetric and broken phases, we study sound velocity effects on the acoustic phase transition gravitational wave spectra in the sound shell model. 
We demonstrate that different sound velocities could obviously modify the peak frequency and peak amplitude of the gravitational wave power spectra. 
Therefore, taking more realistic sound velocities might provide more accurate predictions for various gravitational wave experiments.
\end{abstract}
\maketitle

\section{introduction}
Since the discovery of the gravitational wave (GW) by LIGO and Virgo~\cite{LIGOScientific:2016aoc} and Higgs boson at LHC~\cite{ATLAS:2012yve,CMS:2012qbp}, the cosmological first-order phase transition has attracted a lot of attention, and the corresponding phase transition gravitational wave (PTGW) could provide new perspectives to understand the fundamental problems of particle cosmology, including the baryon asymmetry of the Universe, dark matter formation mechanism, primordial black holes, primordial magnetic field, and spontaneous symmetry breaking in the early Universe. 
Future GW experiments, such as TianQin~\cite{TianQin:2015yph,TianQin:2020hid}, LISA~\cite{LISA:2017pwj}, Taiji~\cite{Hu:2017mde}, etc, may be able to detect the PTGW signals generated by bubble collision, turbulence and sound wave mechanisms~\cite{Hindmarsh:2013xza,Hindmarsh:2015qta,Hindmarsh:2017gnf}. 
For most cases of thermal phase transition, GW signals from the sound wave mechanism  are dominant~\cite{Caprini:2019egz}.

To extract more reliable information of the early Universe from the GW spectra, it is necessary to precisely calculate the PTGW spectra, especially the dominant source, acoustic PTGW spectra from sound wave mechanism.
Currently, there are several methods to calculate the acoustic PTGW.
One is the numerical method directly from lattice hydrodynamic simulation of the coupled fluid-field system~\cite{Hindmarsh:2013xza,Hindmarsh:2015qta,Hindmarsh:2017gnf}
or a simplified numerical method called the hybrid simulation~\cite{Jinno:2020eqg,Jinno:2021ury}.
The other methods depend on some specific models~\cite{Hindmarsh:2016lnk,Hindmarsh:2019phv,Konstandin:2017sat}.
Here, we use the sound shell model (SSM)~\cite{Hindmarsh:2016lnk,Hindmarsh:2019phv} developed by Mark Hindmarsh to study the sound velocity effects on the PTGW spectra.
In the SSM, the calculation of the shear stress unequal time correlator (UETC)\footnote{The shear stress UETC is the source of GW.} is converted to the computation of the velocity UETC since the dominant contribution of shear stress comes from the fluid velocity field. 
And the essential ingredient for the velocity UETC is the velocity profile.
Then the corresponding effect can transfer to the shear stress UETC, and finally the velocity profile would affect the GW power spectra.
Therefore, we should consider more realistic sound velocities in the symmetric and broken phases to obtain more reliable velocity profiles.
Taking sound velocity effects into account, we calculate the velocity power spectra and the corresponding PTGW power spectra in the SSM.

This paper is organized as follows. In Sec.~\ref{SSM:ssm}, we give a brief introduction to the SSM.
Section.~\ref{SSM:PTdynamics} presents the phase transition dynamics and the bubble collision time distribution.
In Sec.~\ref{SSM:profile}, we discuss the self-similar fluid profiles around the expanding bubbles with different sound velocities in symmetric and broken phases.
Then, the GW spectral density from shear stress UETC is reviewed in Sec.~\ref{SSM:GWUETC}.
We study the velocity power spectra in Sec.~\ref{SSM:velocity} and the GW power spectra in Sec.~\ref{SSM:GWspectrum} for different sound velocities.
The discussions and conclusion are given in Secs.~\ref{SSM:dis} and \ref{SSM:con}, respectively.

\section{sound shell model in a nutshell}\label{SSM:ssm}

\begin{figure*}
	\centering
	\includegraphics[width=0.5\textwidth]{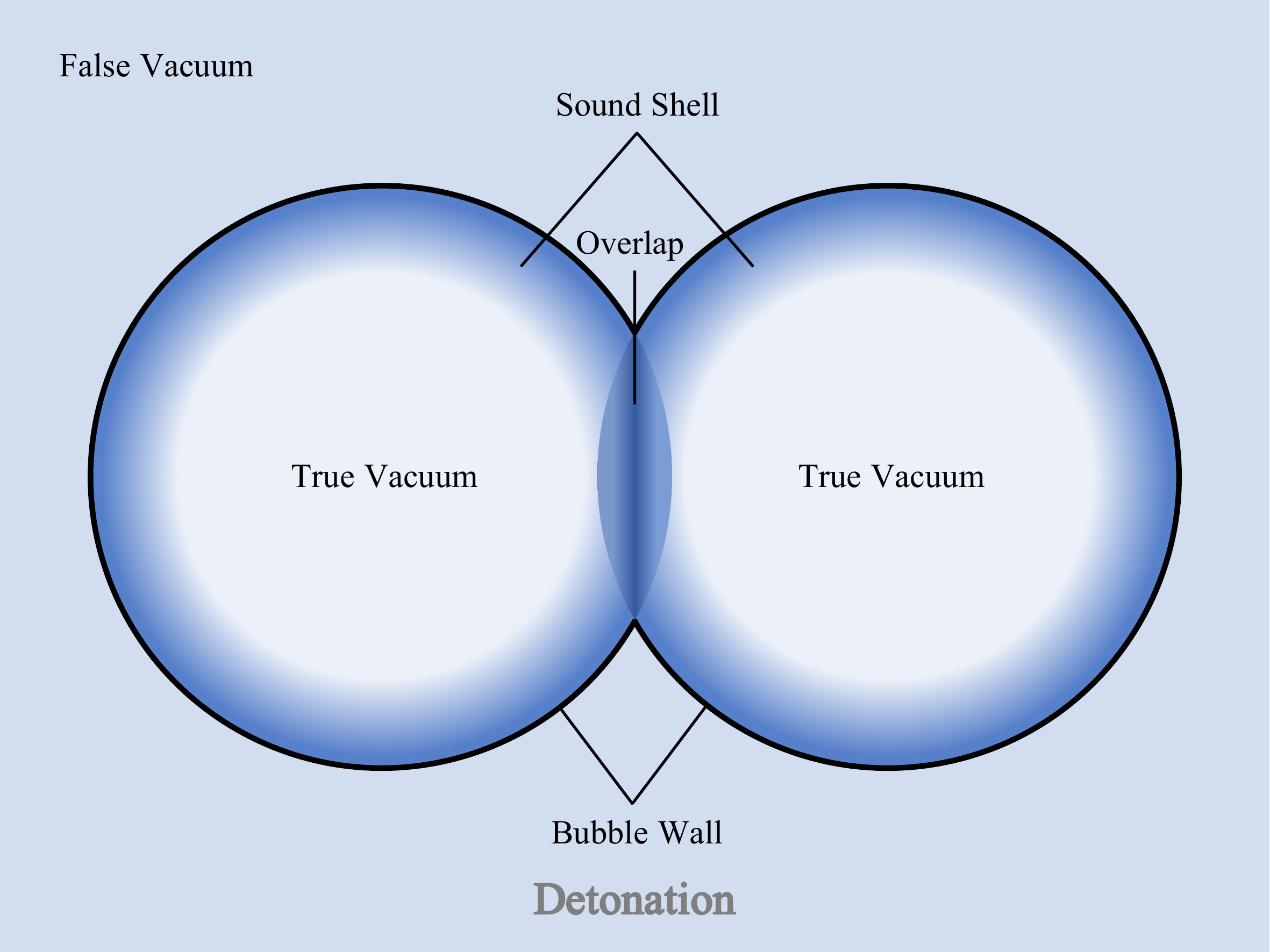}%
	\includegraphics[width=0.5\textwidth]{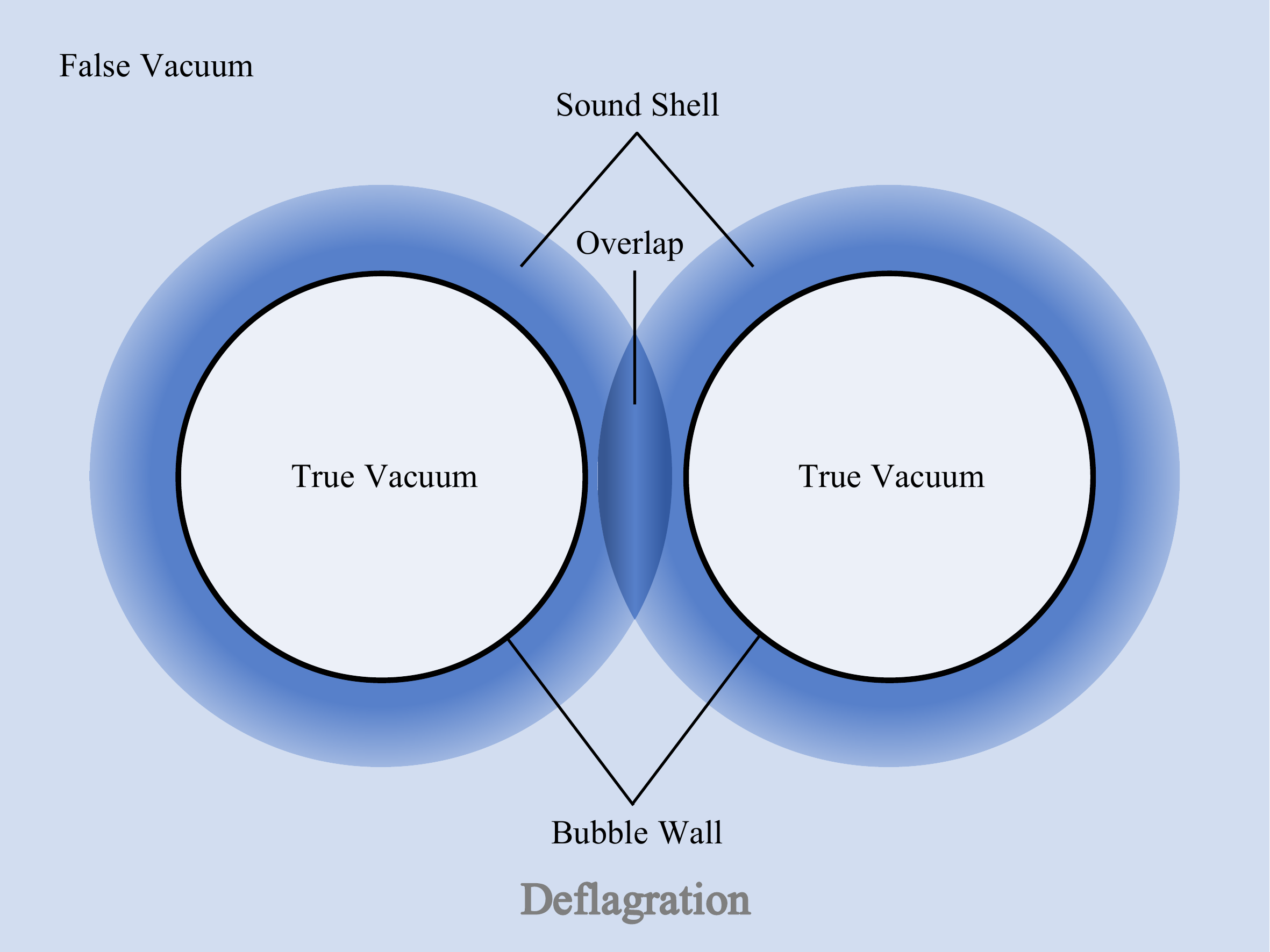}
	\caption{Schematic diagrams of the superposition of self-similar profiles for detonation (left) and deflagration (right) in the SSM. The black circle represents the bubble wall, and the blue shaded ring depicts the sound shell.}
	\label{ssm}
\end{figure*}
Current studies show there are three mechanisms, which are bubble collision, turbulence, and sound waves, for the production of  PTGW. 
In most of thermal phase transitions, the PTGW from sound waves is much stronger than the other two sources.  
The SSM provides a simple approach to obtain the GW prediction without time-consuming lattice simulations.  
This method is based on the  fact that GW is determined by the anisotropic stress tensor $\Pi_{i j}$, and for sound waves, $ \Pi_{i j} \sim\left[\gamma^{2}(\rho+p) v_{i} v_{j}\right]^{T T}$.
Thus, the calculation of GW power spectra from sound waves is converted to the study of velocity power spectra.
In this section, we briefly introduce the basic setup of the SSM~\cite{Hindmarsh:2016lnk,Hindmarsh:2019phv,Cutting:2019zws,Gowling:2021gcy,Giese:2021dnw,Guo:2020grp} with the schematic process shown in Fig.~\ref{ssm}.

In the SSM, a basic approximation is that the source of anisotropic stress is originated from the fluid sound wave.
And the generation of a sound wave is a random process which is initiated after the collision of bubbles.
The GW spectra could be obtained from the shear stress UTEC, which could be calculated from the velocity UETC (or the spectral density of the velocity field).
One basic assumption for the SSM is that fluid velocity field is the superposition of self-similar velocity profiles generated by expanding bubbles.
Here, in Fig.~\ref{ssm}, we illustrate superposition of velocity profiles for two different hydrodynamical modes, which are detonation (left panel) and deflagration (right panel).  
The bubble wall is represented by the black circle, and the blue shaded ring denotes the sound shell.
For calculating the bubble lifetime in the SSM, we define that a bubble is completely destroyed when half of it has merged with other bubbles.
Hence, to obtain more reliable velocity power spectra and PTGW, we should try to give more realistic velocity profiles or improve the modeling of the bubble lifetime.
And in this work, we consider more realistic sound velocities in both symmetric and broken phases to study the sound velocity effects on the velocity profile and the resulting acoustic GW power spectra.

Here, we present the basic steps for the calculation of PTGW in the SSM:
\begin{itemize}
	\item Derive the self-similar velocity and enthalpy profiles of a single expanding bubble based on the specific equation of sate (EoS) which can incorporate different sound velocities in the symmetric and broken phases. 
	\item Obtain the single-bubble plane wave amplitude from single-bubble self-similar profile. Then derive the plane wave amplitude correlation function for the velocity field generated by $N$ randomly placed bubbles in a given volume. 
	\item Estimate the bubble collision time distribution for a specific nucleation history, and derive the velocity spectral density and power spectrum.
	\item Finally derive the GW power spectrum using the relation between the shear stress UETC and the velocity spectral density.
\end{itemize}

\section{phase transition dynamics and bubble collision time distribution}\label{SSM:PTdynamics}

For a cosmological first-order phase transition, the starting point of the phase transition dynamics is the bubble nucleation rate per unit time per unit volume
\begin{equation}
	\Gamma(t) = \Gamma_0e^{-S_E}\,\,,
\end{equation}
where $S_E$ is the bounce action and can be obtained by solving the equation of motion of order-parameter field with a given effective potential~\cite{Coleman:1977py,Callan:1977pt,Linde:1980tt,Linde:1981zj,Wang:2020jrd}. 
However, to calculate the PTGW and other phenomenology, we need more parameters: the characteristic temperature $T_*$, the characteristic length scale $R_*$, the energy budget, and the bubble wall velocity $v_w$.
In a first-order phase transition, $T_*$ is the temperature at which GW is generated.
And this temperature is conventionally approximated as nucleation temperature or percolation temperature in most of the literature.
To derive the percolation temperature, the fraction of space remaining in the symmetric phase (i.e., the probability of finding a point still in the false vacuum) is important, and it is
\begin{equation}
	P(t) = \exp\left(-\frac{4\pi}{3}\int_{t_c}^{t}dt'\Gamma(t')R(t',t)\right)\,\,,
	\label{eq:Pt}
\end{equation}
where the radius of the bubble nucleated at $t'$ is $R(t',t) =\int_{t'}^{t} dt''v_w = v_w(t -t')$ neglecting the expansion of the Universe.
The overlap of the bubbles is taken into account by the exponentiation of Eq.~\eqref{eq:Pt}.
\begin{figure}
	\centering
	\includegraphics[width=0.5\textwidth]{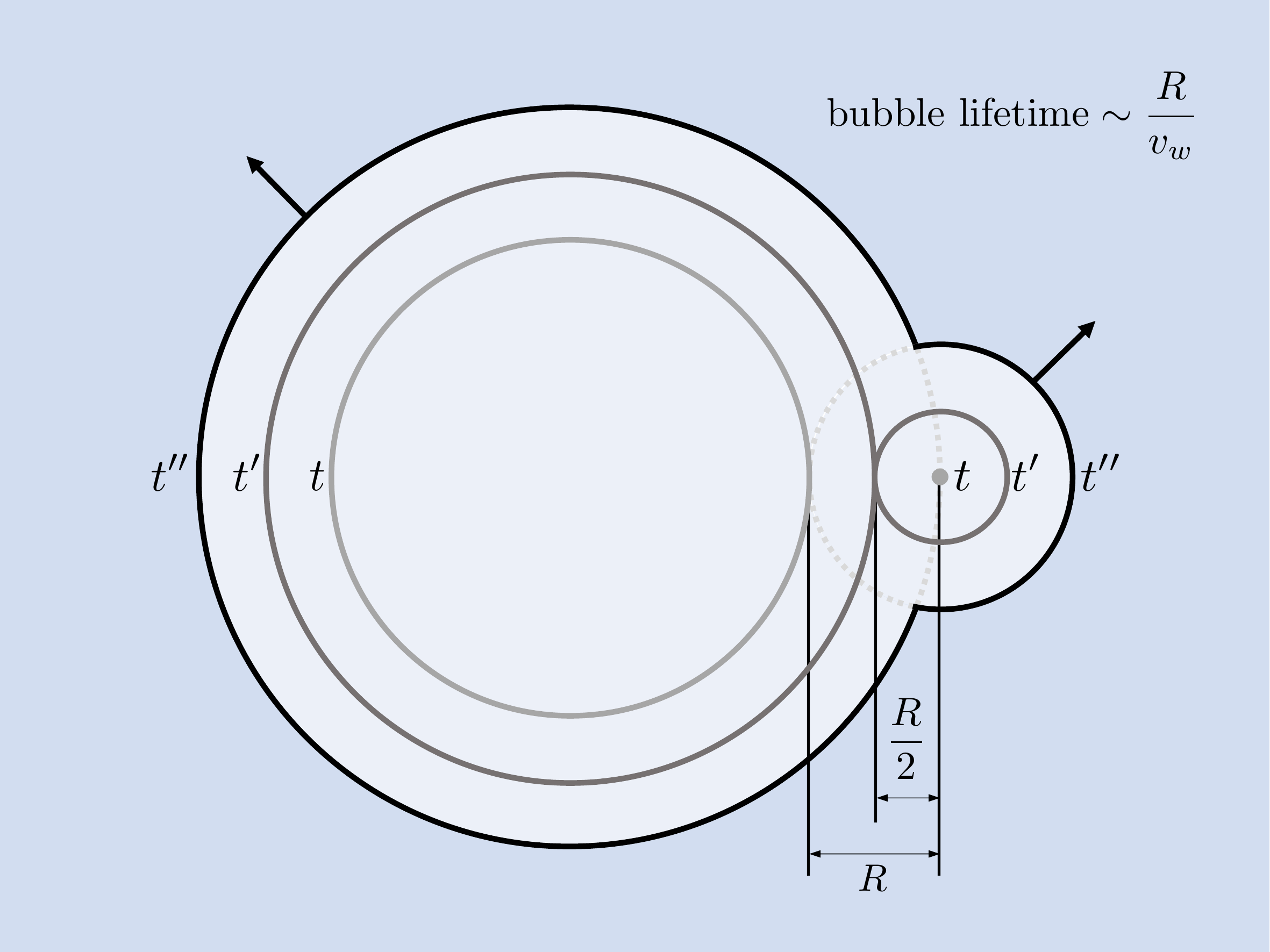}
	\caption{Schematic diagram for the bubble lifetime. At time $t$, there is a bubble just formed at the light gray dot and an expanding bubble depicted by the light gray circle. The distance between the light gray dot and the wall of expanding bubble is $R$. At $t'$, the two bubbles firstly collide and start to merge, and $t'-t = R/(2v_w)$. At $t''$, half of the small bubble has merged with the large bubble, and we define that the small bubble disappears. Hence, the lifetime of the small bubble is $R/v_w$.}
	\label{lifetime}
\end{figure}
Usually, the  characteristic length scale is chosen as the mean bubble separation $R_*$ when GWs are produced.
And it can be conventionally expressed as 
\begin{equation}
R_* = \frac{(8\pi)^{1/3}}{\beta}v_w,\quad \beta = HT\frac{dS_E}{dT}\Big|_{T = T_*}\,\,.
\end{equation}
Then the bubble number density is $n_b = R_*^{-3}$. 
The energy budget and the bubble wall velocity are strongly related, and they can significantly affect the strength of  PTGW.
And the energy budget is usually obtained by a model-independent method~\cite{Espinosa:2010hh,Leitao:2014pda,Giese:2020rtr,Giese:2020rtr,Wang:2020nzm}, which is developed with some specific models of EoS. 
The bubble wall velocity $v_w$ is conventionally chosen as a free parameter in the relevant studies of PTGW and electroweak baryogenesis.
However, it should be determined by the interaction between the bubble wall and the surrounding plasma~\cite{Moore:1995ua,Moore:1995si,Moore:2000wx,John:2000zq,Konstandin:2014zta,Kozaczuk:2015owa,Hoche:2020ysm,Wang:2020zlf,Dorsch:2021nje,Dorsch:2021ubz,Laurent:2020gpg,Azatov:2020ufh,Gouttenoire:2021kjv}.

To derive the GW power spectrum in the SSM, one of the essential quantities is the bubble collision time distribution.
And for the calculation of bubble collision time distribution, the key parameters are the bubble nucleation rate $\Gamma(t)$, the fraction of space remaining in the symmetric phase $P(t)$, the area of the bubble wall (or the phase boundary) per unit volume $\mathcal{A}(t)$, and the bubble wall velocity $v_w$.
In Fig.~\ref{lifetime}, we demonstrate the bubble lifetime schematically.
At time $t$, there is a bubble just formed at the light gray dot, and an expanding bubble depicted by light gray circle. The distance between the light gray dot and the wall of expanding bubble is $R$. At $t'$, the two bubbles first collide and start to merge, and $t'-t = R/(2v_w)$. At $t''$, half of the small bubble has merged with the large bubble, we define that the small bubble disappears. Hence, the lifetime of the small bubble is $R/v_w$.
So in time interval $dt$, the bubbles nucleated in the volume $\mathcal{A}(t + R/v_w)dR$ are all destroyed.
Hence we have
\begin{equation}
\begin{split}
&d^2n_b = \left[\mathcal{A}(t + R/v_w)dR\right]\left[\Gamma(t)dt\right]\,\,,\\
&d\left(\frac{dn_b}{dR}\right)  = \mathcal{A}(t + R/v_w)\Gamma(t)dt\,\,.
\end{split}
\end{equation}
Then the bubble size distribution can be derived as
\begin{equation}
\frac{dn_b}{dR} = \int_{t_c}^{t'}\mathcal{A}(t + R/v_w)\Gamma(t)dt\,\,,
\end{equation}
where $t_c$ is the time at which the effective potential has degenerate minima and the area per unit volume of the bubble wall is
\begin{equation}
\mathcal{A} = -\frac{1}{v_w}\frac{d P}{dt}\,\,.
\end{equation}
Here, we introduce the probability density distribution of lifetime $n(\mathcal{T}_i)$, which will be described in detail in Sec.~\ref{SSM:velocity}.
The relation between the probability density distribution of lifetime $n(\mathcal{T}_i)$ and the bubble number density $n_b$ is 
\begin{equation}
\int n(\mathcal{T}_i)d\mathcal{T}_i = n_b,\quad n(\mathcal{T}_i)d\mathcal{T}_i = dn_b\,\,.
\end{equation}
Hence,
\begin{equation}
\frac{dn_b}{dR} = n(\mathcal{T}_i)\frac{d\mathcal{T}_i}{dR},\quad d\mathcal{T}_i = dR/v_w\,\,.
\end{equation}
We can derive
\begin{equation}
	\frac{\beta}{R_*^3}f_{\rm col}(\tilde{\mathcal{T}})d\mathcal{T}_i = n(\mathcal{T}_i)d\mathcal{T}_i\,\,;
	\label{eq:coltd1}
\end{equation}
then, the collision time distribution can be defined as
\begin{equation}
f_{\rm col}(\tilde{\mathcal{T}}) = v_w\frac{R_*^3}{\beta}\frac{dn_b}{dR}\,\,.
\end{equation}
Here, $\mathcal{T}_i$ is the lifetime of bubbles, and $\tilde{\mathcal{T}} = \beta\mathcal{T}_i$. In this work, we only consider the exponential nucleation.
Taking a suitable approximation, one can derive the collision time distribution as follows~\cite{Hindmarsh:2019phv}
\begin{equation}
f_{\rm col}(\tilde{\mathcal{T}}) = e^{-\tilde{\mathcal{T}}}\,\,.
\label{eq:coltd2}
\end{equation}

\section{Self-similar fluid shell with realistic sound velocity}\label{SSM:profile}
\begin{figure*}[t]
	\centering
	\includegraphics[width=0.5\textwidth]{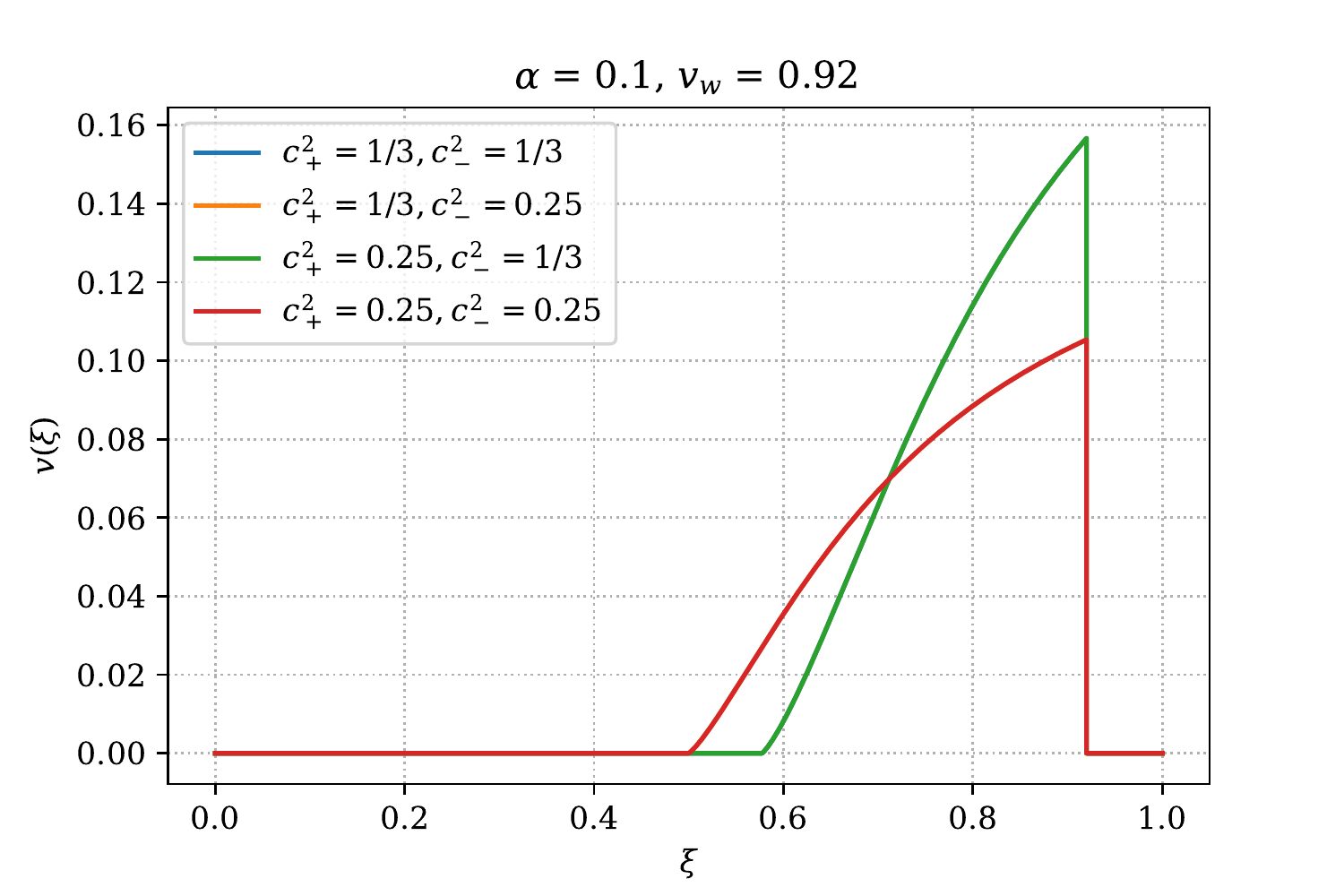}%
	\includegraphics[width=0.5\textwidth]{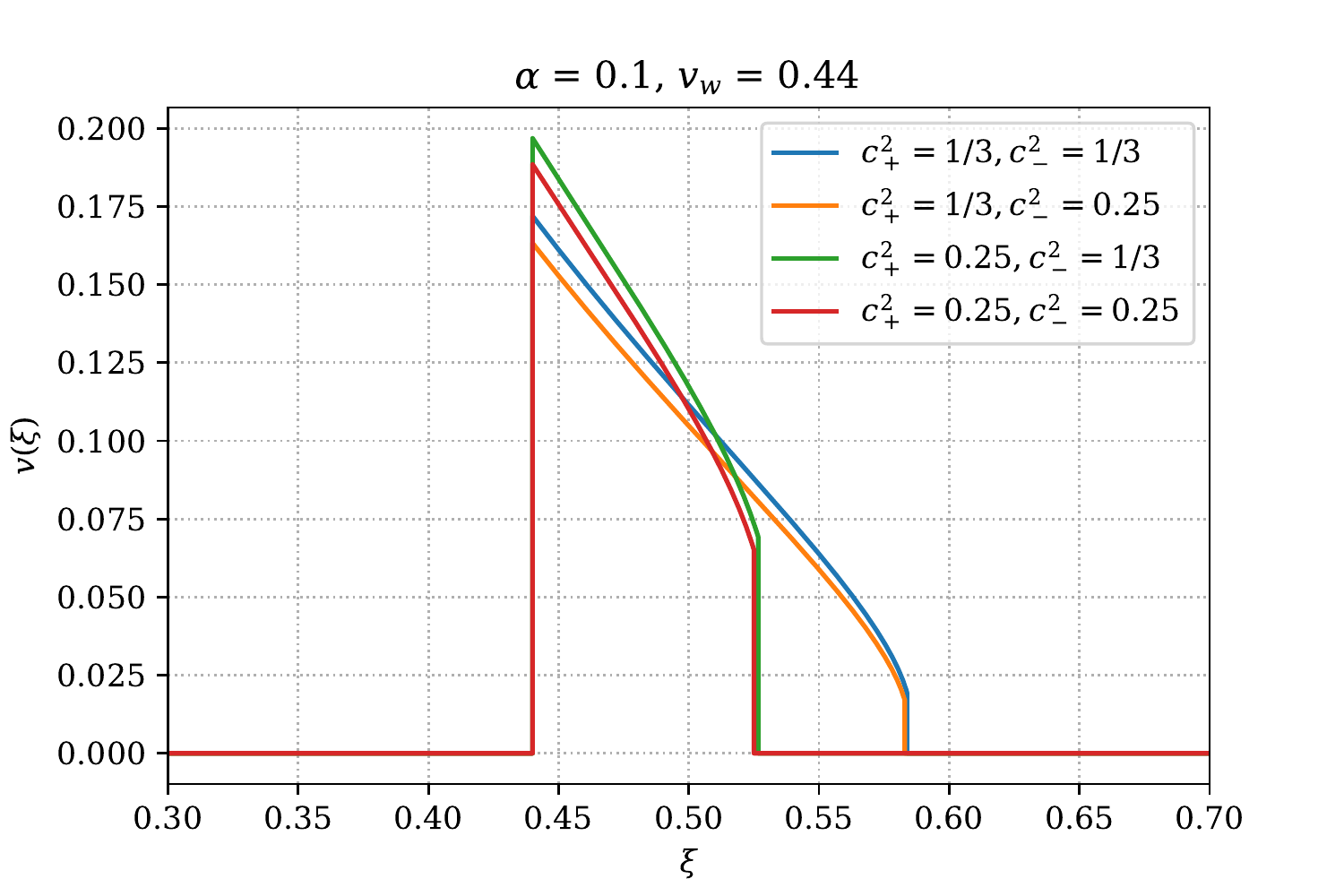}
	
	\includegraphics[width=0.5\textwidth]{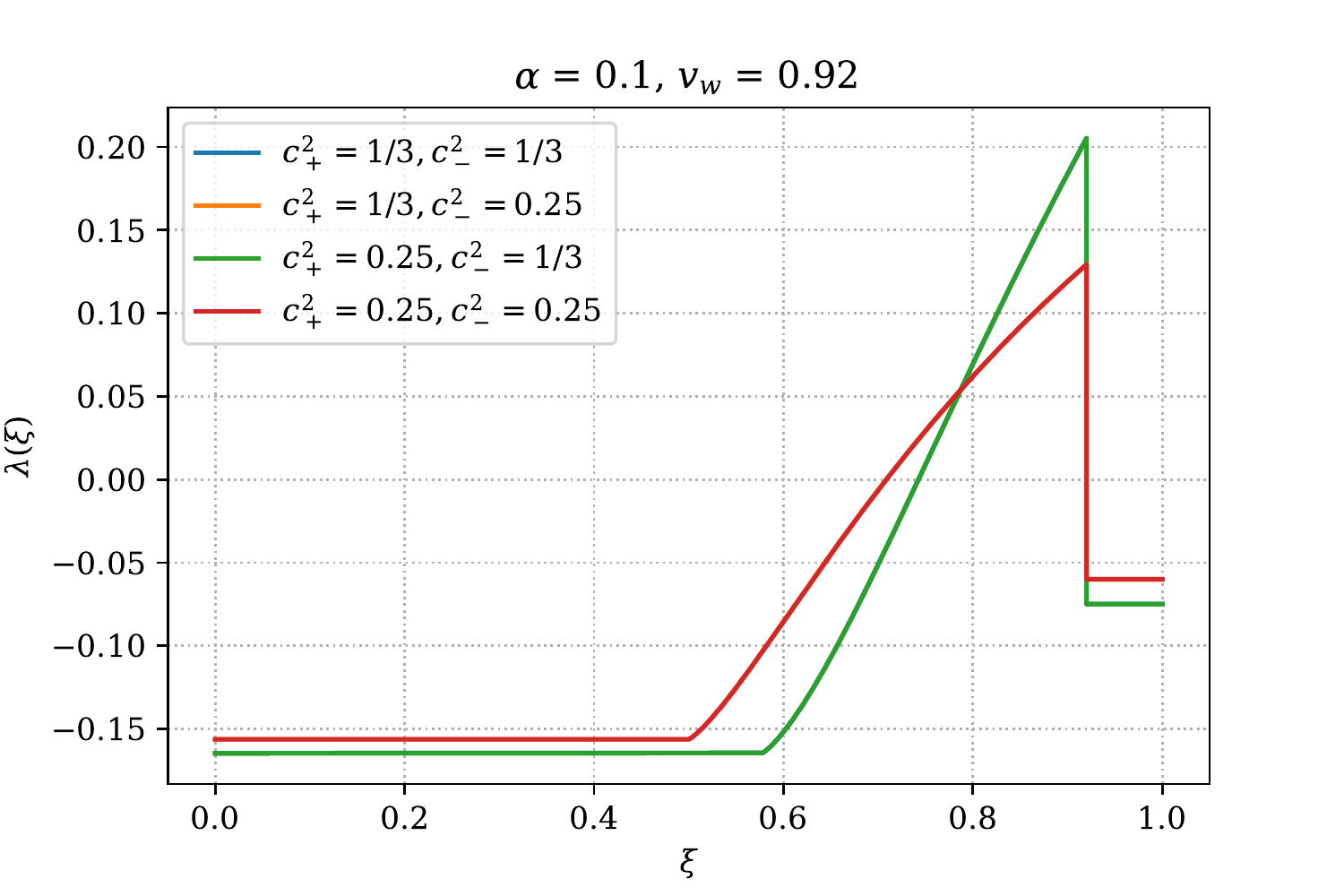}%
	\includegraphics[width=0.5\textwidth]{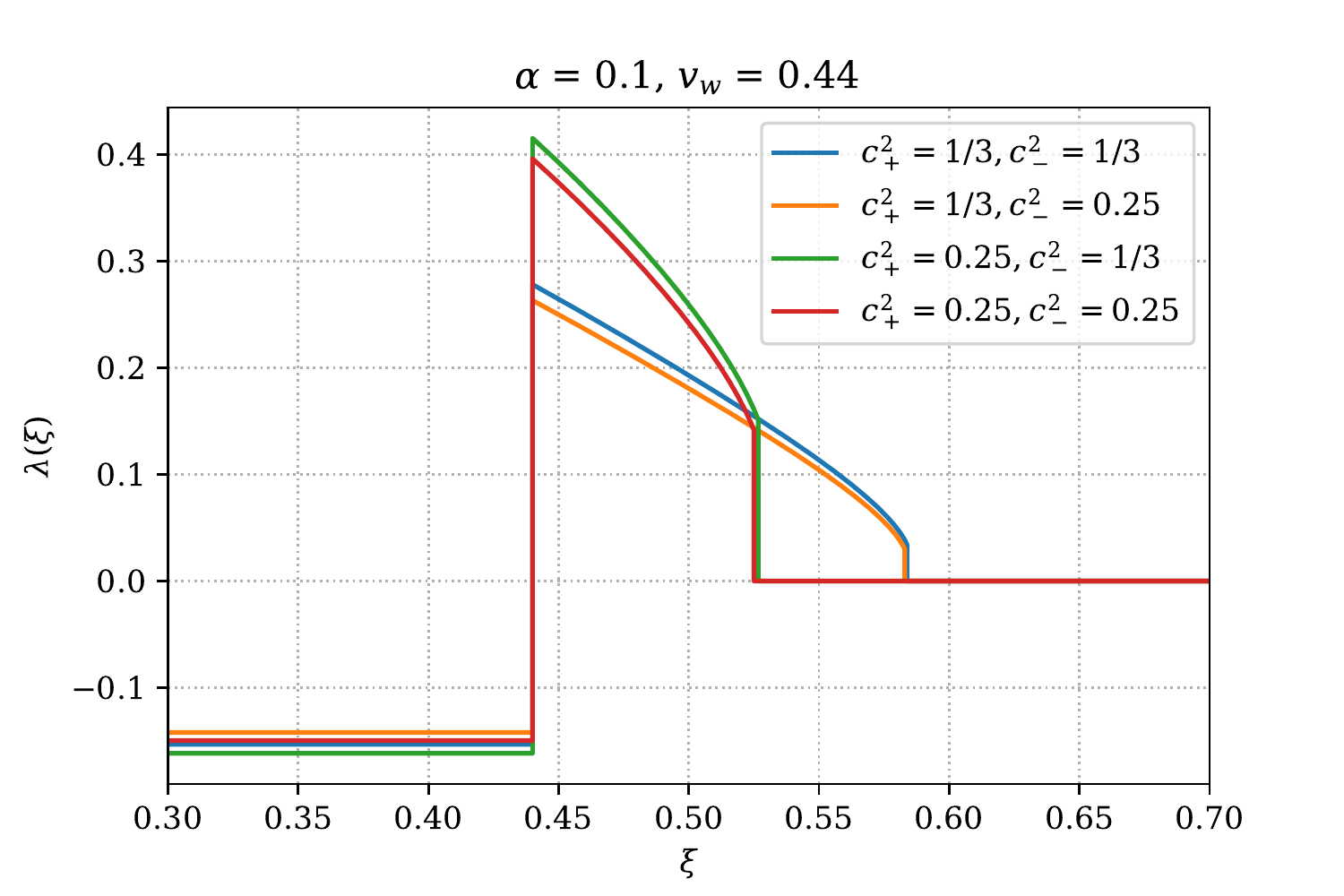}
	
	\caption{The velocity and energy fluctuation profiles of the deflagration mode and detonation mode for the DSVM of EoS. The left column is detonation, and the right column is deflagration. Different colors represent different combinations of sound velocities in symmetric and broken phases.}\label{Fig:profile}
\end{figure*}

In this work, we consider the sound velocity effects on PTGW power spectra. 
According to the basic setup of the SSM, the effect of sound velocity is originated from the self-similar profiles.
To derive the sound-velocity-dependent profiles, we use different sound velocities model (DSVM)~\cite{Leitao:2014pda,Giese:2020rtr,Giese:2020znk,Wang:2020nzm} as the EoS to 
derive the initial self-similar fluid profiles. 
In the DSVM, the EoS is
\begin{equation}
	\begin{split}
		p_+ = c_+^2a_+T_+^4 - \epsilon,&\quad e_+ = a_+T_+^4 + \epsilon\,\,,\\
		p_- = c_-^2a_-T_-^4 ,&\quad e_- = a_-T_-^4\,\,,\label{constantmodel}
	\end{split}
\end{equation}
where $a_{\pm} = g_{\pm}\pi^2/30$, and $g_{\pm}$ are the degree of freedom for the symmetric and broken phases respectively ($+$ for symmetric phase and $-$ for broken phase).
To obtain the initial condition of the SSM, we need to solve the hydrodynamical equations~\cite{Kurki-Suonio:1984zeb,Kamionkowski:1993fg,Espinosa:2010hh,Leitao:2010yw,Leitao:2014pda,Giese:2020rtr,Giese:2020znk,Wang:2020nzm}
\begin{equation}
	\begin{split}
		2\frac{v}{\xi} &= \gamma^2(1 - v\xi)\left[\frac{\mu^2}{c_s^2} - 1\right]\partial_\xi v\,\,,\\
		\frac{\partial_\xi w}{w} &= \left(1 + \frac{1}{c_s^2}\right)\mu\gamma^2\partial_\xi v\,\,,\\
	\end{split}
\end{equation}
where
\begin{equation}
	\mu(\xi,v) = \frac{\xi - v}{1 - \xi v}\,\,.
\end{equation}
With different boundary conditions, we can derive the velocity and enthalpy profiles for three stable hydrodynamical modes, which are deflagration, hybrid, and detonation.
In the DSVM of EoS, the strength parameter can be defined as 
\begin{equation}
	\alpha = \frac{1}{3}\left[\frac{1 - c_+^2/c_-^2}{1 + c_+^2} + \frac{(1 + 1/c_-^2)\epsilon}{w_+}\right]\,\,.
\end{equation}
One can define the the energy fluctuation variable as
\begin{equation}
	\lambda(x) = \frac{e(x) - \bar{e}}{\bar{w}}\,\,,
\end{equation}
where $\bar{w}$ and $\bar{e}$ are mean enthalpy and mean energy density, respectively.
Then, according to the EoS and the fluid profiles, for detonation we have 
\begin{equation}
	\begin{split}
		\lambda_{\rm DT}(\xi) =& \left[\frac{cw(\xi)/\bar{w}}{(1+d(3\alpha - b)/a)} - 1\right]\\
		&\times \left[\frac{1}{d} + \frac{3\alpha - b}{a}\right]\\
		=&\frac{1}{1 + c_-^2}\left[\frac{w(\xi)}{\bar{w}} - (1 + 3c_-^2\alpha)\right]\,\,,
	\end{split}
	\label{eq:DTlambda}
\end{equation}
and for deflagration
\begin{equation}
	\begin{split}
		\lambda_{\rm DF}(\xi) =& \left[\frac{w(\xi)/\bar{w} + d(3\alpha - b)/a}{1 + d(3\alpha - b)/a} - 1\right]\\
		&\times \left[\frac{1}{d} + \frac{3\alpha - b}{a}\right]\,\,,
	\end{split}
\end{equation}
where
\begin{equation}
	\begin{split}
		a = 1 + 1/c_-^{2}\,\,,&\quad b = \frac{1 - c_+^2/c_-^2}{1 + c_+^2}\,\,,\\
		c = \frac{1 + c_+^2}{1 + c_-^2}\,\,,&\quad d = 1 + c_+^2\,\,.\\
	\end{split}
\end{equation}
In this work, we do not consider the hybrid mode.  We exemplify the corresponding velocity and energy fluctuation profiles of the deflagration and detonation modes for the DSVM of the EoS in Fig.~\ref{Fig:profile}.
The left column is detonation, and the right column is deflagration. Different colors represent different combinations of sound velocities in symmetric and broken phases.  
From the numerical results shown in Fig.~\ref{Fig:profile}, we can see that the sound velocities in symmetric and broken phases have obvious impact on the velocity and energy fluctuation profiles.
According to Fig.~\ref{Fig:profile} and Eq.~\eqref{eq:DTlambda}, we can find only the sound velocity of the broken phase can affect the corresponding profiles of detonation.
For a given phase transition model,  precise calculation of the velocity and energy profiles  is crucial to the precise prediction of PTGW spectra since they act as the initial conditions for the calculation of GW spectra in the SSM.

\section{Gravitational wave spectral density from shear stress unequal time correlator}\label{SSM:GWUETC}

For most of the cosmological first-order phase transition, the phase transition duration is much shorter than the Hubble time.
Thus, we could ignore the cosmic expansion if the supercooling is not too strong.
In the weak field approximation, we have the following  metric perturbation: 
\begin{equation}
	d s^{2}=-d t^{2}+\left(\delta_{i j}+h_{i j}\right) d x^{i} d x^{j}\,\,.
\end{equation}
Substituting the above perturbation into the Einstein field equation, we obtain the GW equation for
the transverse-traceless part of the energy momentum tensor:
\begin{equation}
	\ddot{h}_{i j}-\nabla^{2} h_{i j}=16 \pi G \Pi_{i j}\,\,.
\end{equation}
We can write it in a more convenient form,
\begin{equation}
	\ddot{\psi}_{i j}-\nabla^{2} \psi_{i j}=16 \pi G \tau_{i j}\,\,,
	\label{gwsourceeq}
\end{equation}
with  $h_{i j}=\psi_{i j}^{T T}$ and $\Pi_{i j}=\tau_{i j}^{TT}$.
Using the projection operator,  we can derive 
\begin{equation}
	\dot{h}_{i j}(\mathbf{k}, t)=\lambda_{i j, k l}(\mathbf{k}) \dot{\psi}_{k l}(\mathbf{k}, t)\,\,,
\end{equation}
where $\lambda_{i j, k l}(\mathbf{k})=P_{i k}(\mathbf{k}) P_{j l}(\mathbf{k})-\frac{1}{2} P_{i j}(\mathbf{k}) P_{k l}(\mathbf{k})$ and $P_{i j}(\mathbf{k})=\delta_{i j}-\hat{k}_{i} \hat{k}_{j}$.
Ignoring the scalar field contribution to the energy-momentum tensor, we have
\begin{equation}
	\tau_{i j}=(e+p) \gamma^{2} v_{i} v_{j}\,\,.
	\label{Eq:tauij}
\end{equation}
The solution is 
\begin{equation}
	\psi_{i j}(\mathbf{k}, t)=16 \pi G \int_{0}^{t} d t^{\prime} \frac{\sin \left[k\left(t-t^{\prime}\right)\right]}{k} \tau_{i j}\left(\mathbf{k}, t^{\prime}\right)\,\,.
	\label{eq:EEsol}
\end{equation}
Therefore, we could obtain the GW energy density 
\begin{equation}
	\rho_{\mathrm{gw}}=\frac{1}{32 \pi G} 	\left\langle  \dot{h}_{i j}(x) \dot{h}_{ij}(x) \right\rangle \,\,.
\end{equation}
And in Fourier space, we could define the spectral density of $\dot{h}$ as
\begin{equation}
	\left\langle\dot{h}_{i j}(\mathbf{k}, t) \dot{h}_{i j}^{*}\left(\mathbf{k}^{\prime}, t\right)\right\rangle=P_{\dot{h}}(\mathbf{k}) \delta^{3}\left(\mathbf{k}-\mathbf{k}^{\prime}\right)\,\,.
	\label{eq:Phdot1}
\end{equation}
Then, we have the following expression for the GW energy density: 
\begin{equation}
	\rho_{\mathrm{gw}}=\frac{1}{32 \pi G} \int d^{3} k P_{\dot{h}}(k)=\frac{1}{32 \pi G} \frac{1}{2 \pi^{2}} \int d k k^{2} P_{\dot{h}}(k)\,\,.
\end{equation}
For GW experiments, it is also convenient to define the so-called power spectrum, which is easier to compare with the noise of GW detectors.
Here, we give the power spectrum
\begin{equation}
	\mathcal{P}_{\dot{h}}=\frac{k^{3}}{2 \pi^{2}} P_{\dot{h}}(k)\,\,;
\end{equation}
then,
\begin{equation}
	\rho_{\mathrm{gw}}=\frac{1}{32 \pi G} \int \frac{d k}{k} \mathcal{P}_{\dot{h}}(k)\,\,.
\end{equation}

In cosmology, it is common to use the energy density $\Omega_{\mathrm{gw}}=\rho_{\mathrm{gw}} / \rho_{\rm crit}$,
where $\rho_{\rm crit} = 3H_0^2/(8\pi G)$ is the critical density, and $H_0$ is the current Hubble rate.
Finally, we could define the  GW power spectrum as 
\begin{equation}\label{pgl}
	\mathcal{P}_{\mathrm{gw}}(k) \equiv \frac{d \Omega_{\mathrm{gw}}}{d \ln (k)}=\frac{1}{\rho_{\rm crit}} \frac{1}{32 \pi G} \mathcal{P}_{\dot{h}}(k)=\frac{1}{12 H^{2}} \mathcal{P}_{\dot{h}}(k)\,\,.
\end{equation}
Hence, to derive the GW power spectrum, we need to first obtain the spectral density.
Substituting Eq.~\eqref{eq:EEsol} into Eq.~\eqref{eq:Phdot1}, we can obtain
\begin{equation}
\begin{aligned}
\left\langle\dot{h}_{\mathbf{k}}^{i j}(t) \dot{h}_{\mathbf{k}^{\prime}}^{i j}(t)\right\rangle=& (16 \pi G)^{2} \int_{0}^{t} d t_{1} d t_{2} \cos \left[k\left(t-t_{1}\right)\right] \cos \left[k\left(t-t_{2}\right)\right]\\
&\times\lambda_{i j, k l}(\mathbf{k})\left\langle\tau^{i j}\left(\mathbf{k}, t_{1}\right) \tau^{k l}\left(\mathbf{k}^{\prime}, t_{2}\right)\right\rangle .
\end{aligned}
\end{equation}
We define the UETC for shear stress as follows:
\begin{equation}\label{suetc}
	\lambda_{i j, k l}(\mathbf{k})\left\langle\tau^{i j}\left(\mathbf{k}, t_{1}\right) \tau^{k l}\left(\mathbf{k}^{\prime}, t_{2}\right)\right\rangle=U_{\Pi}\left(k, t_{1}, t_{2}\right) \delta^{3}\left(\mathbf{k}+\mathbf{k}^{\prime}\right)\,\,.
\end{equation}
Then, the GW spectral density can be obtained as
\begin{equation}\label{pmd}
	P_{\dot{h}}(k, t)=(16 \pi G)^{2} \frac{1}{2} \int_{0}^{t} d t_{1} d t_{2} \cos \left[k\left(t_{1}-t_{2}\right)\right] U_{\Pi}\left(k, t_{1}, t_{2}\right)\,\,.
\end{equation}
Therefore, to calculate the GW power density, we need to calculate the shear stress UETC, which could be obtained by
the velocity UETC or the velocity spectral density as discussed in the following.

\section{velocity field in the Sound Shell Model}\label{SSM:velocity}
In the SSM, the fluid is assumed to be the only source of shear stress.
For nonrelativistic fluid velocities,
based on Eq.~\eqref{Eq:tauij},
we have
\begin{equation}
\tau_{ij} \simeq \bar{w}v_iv_j\,\,.
\end{equation}
From this equation, we could find that the GW generated by sound wave is fully determined by the behavior of the sound shell velocity.
According to energy-momentum conservation, we can derive the linearized fluid equation as
\begin{equation}
\begin{split}
\frac{\dot{e}}{w} + \partial_jv^j &= 0\,\,,\\
\dot{v}^i + \frac{\partial^ip}{w} &= 0\,\,.
\end{split}
\end{equation}
In Fourier space, we have
\begin{equation}
\begin{split}
\dot{\tilde{\lambda}}_{\textbf{q}} + iq_j\tilde{v}^j_{\textbf{q}} &= 0\,\,,\\
\dot{\tilde{v}}_{\bf q}^i + c_s^2iq^i\tilde{\lambda}_{\bf q} &= 0\,\,.
\end{split}
\end{equation}
The solution for the velocity field can be expressed as
\begin{equation}
v^i(\mathbf{x}, t) = \int\frac{d^3q}{(2\pi)^3}\left(v_{\bf q}^ie^{-i\omega t + i\bf q\cdot x } + v_{\bf q}^{*i}e^{i\omega t - i\bf q\cdot x }\right)\,\,,
\end{equation}
and the plane wave amplitude at time $t_i$ is
\begin{equation}
v_{\bf q}^i = \frac{1}{2}\left(\tilde{v}^i(\mathbf{q}, t_i) + \frac{i}{\omega}\dot{\tilde{v}}^i(\mathbf{q},t_i)\right)e^{i\omega t_i}\,\,.
\end{equation}
where $\omega = c_sq$.

In the SSM, the velocity field is the superposition of self-similar velocity profiles which are generated by expanding bubbles.
Hence the velocity field 
\begin{equation}
v_i(\textbf{x},t) = \sum_{n=1}^{N}v_i^n(\textbf{x},t),\quad v_i^n(\textbf{x},t) = \frac{R_i^n}{R^n}v(\xi)\,\,,
\end{equation}
where the index $n$ represent the $n$th bubble, $R_i^n = x_i - x_i^n$ is the bubble radius, $T^n = t - t^n$ is the duration since the nucleation time $t^n$, and $\xi = R^n/T^n$. In Fourier space, the velocity field is
\begin{equation}
\begin{split}
\tilde{v}_i^n(\textbf{q},t) &= \int d^3xv_i^n(\textbf{x},t) e^{-i\textbf{q}\cdot\textbf{x}^n} \\
&= e^{-i\textbf{q}\cdot\textbf{x}^n}i(\mathcal{T}^n)^3\frac{\partial}{\partial z_i}\left(\int d^3\xi\frac{1}{\xi}v(\xi)e^{-iz_i\xi_i}\right)\\
&= e^{-i\textbf{q}\cdot\textbf{x}^n}i(\mathcal{T}^n)^3\hat{z}_i f'(z)\,\,,
\end{split}
\end{equation}
where $z_i = q_i\mathcal{T}^n$.
The function $f(z)$ is defined by angular integration,
\begin{equation}
\begin{split}
f(z) &= \int d^3\xi\frac{1}{\xi}v(\xi)e^{-iz^i\xi^i} \\
&= \frac{4\pi}{z}\int_{0}^{\infty}d\xi v(\xi)\sin(z\xi)\,\,.
\end{split}
\end{equation}
The Fourier-transformed energy fluctuation is 
\begin{equation}
\tilde{\lambda}_n(\textbf{q},t) = e^{-i\textbf{q}\cdot\textbf{x}^n}(\mathcal{T}^n)^3l(z)\,\,,
\end{equation}
where
\begin{equation}
l(z) = \frac{4\pi}{z}\int_{0}^{\infty}d\xi\lambda(\xi)\xi\sin(z\xi)\,\,.
\end{equation}
The plane wave amplitude is
\begin{equation}
v_{\textbf{q},i}^n = i(\mathcal{T}_i^n)^3\hat{z}_ie^{i\omega t_i - i\textbf{q}\cdot\textbf{x}^n}A(z)\,\,,
\end{equation}
where 
\begin{equation}
A(z) = \frac{1}{2}\left[f'(z) + ic_sl(z)\right]\,\,,
\end{equation}
and $\mathcal{T}_i^n = t_i^n - t^n$ is the lifetime of the $n$th bubble.
The velocity field should freely propagate in the broken phase; hence, we use the sound velocity of broken phase $c_-$ to calculate the corresponding properties.
\begin{figure*}[t]
	\centering
	\includegraphics[width=0.5\textwidth]{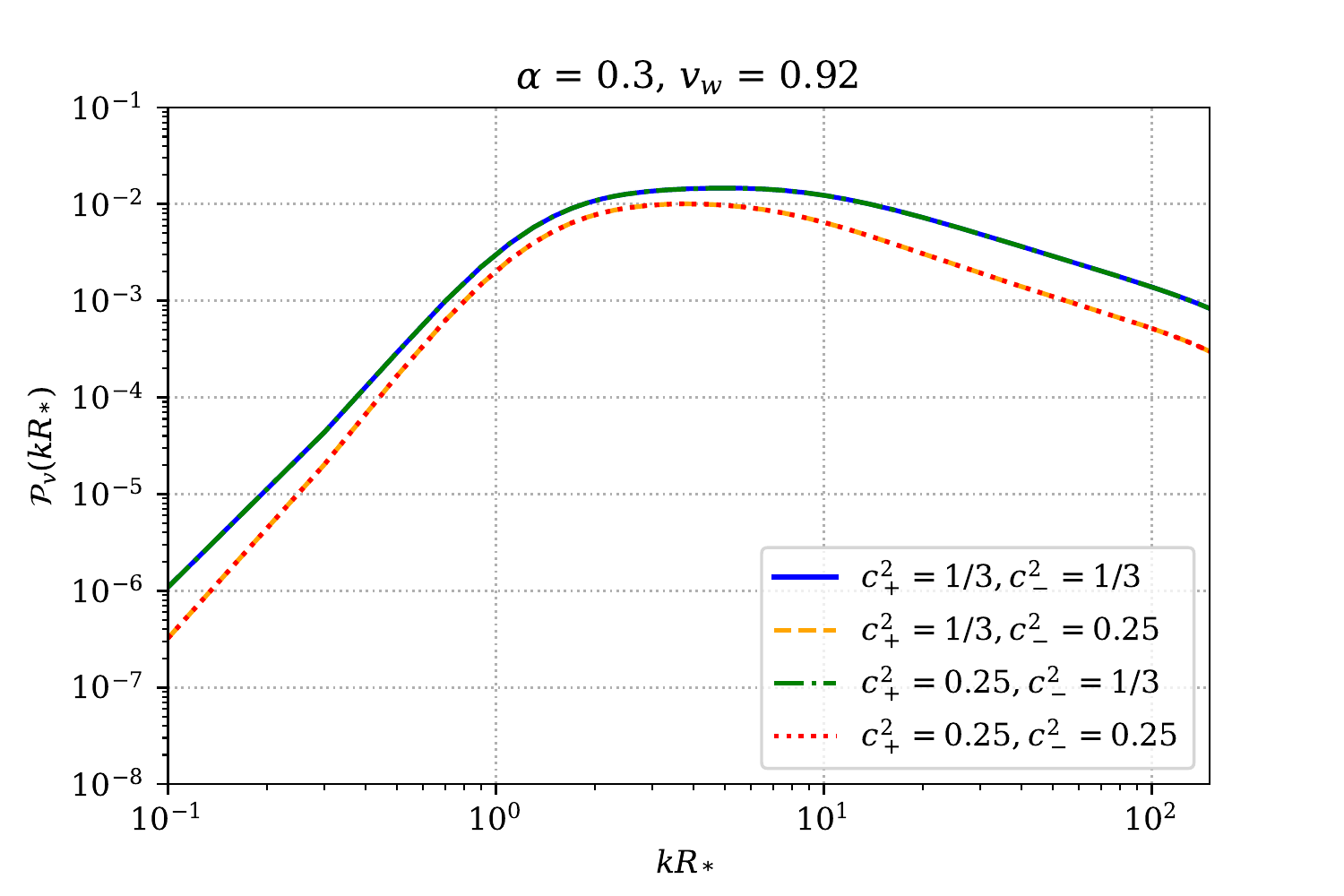}%
	\includegraphics[width=0.5\textwidth]{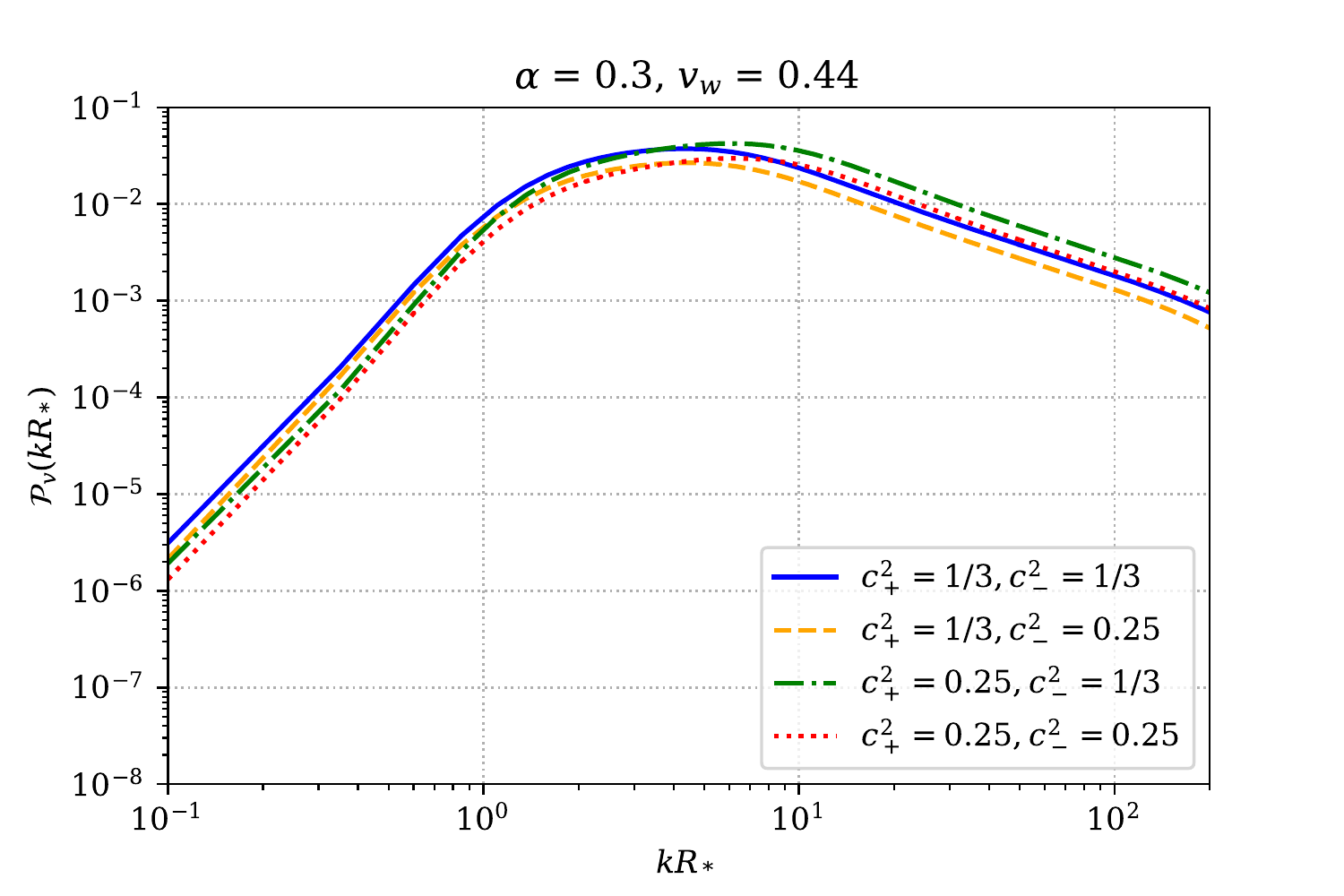}
	
	\includegraphics[width=0.5\textwidth]{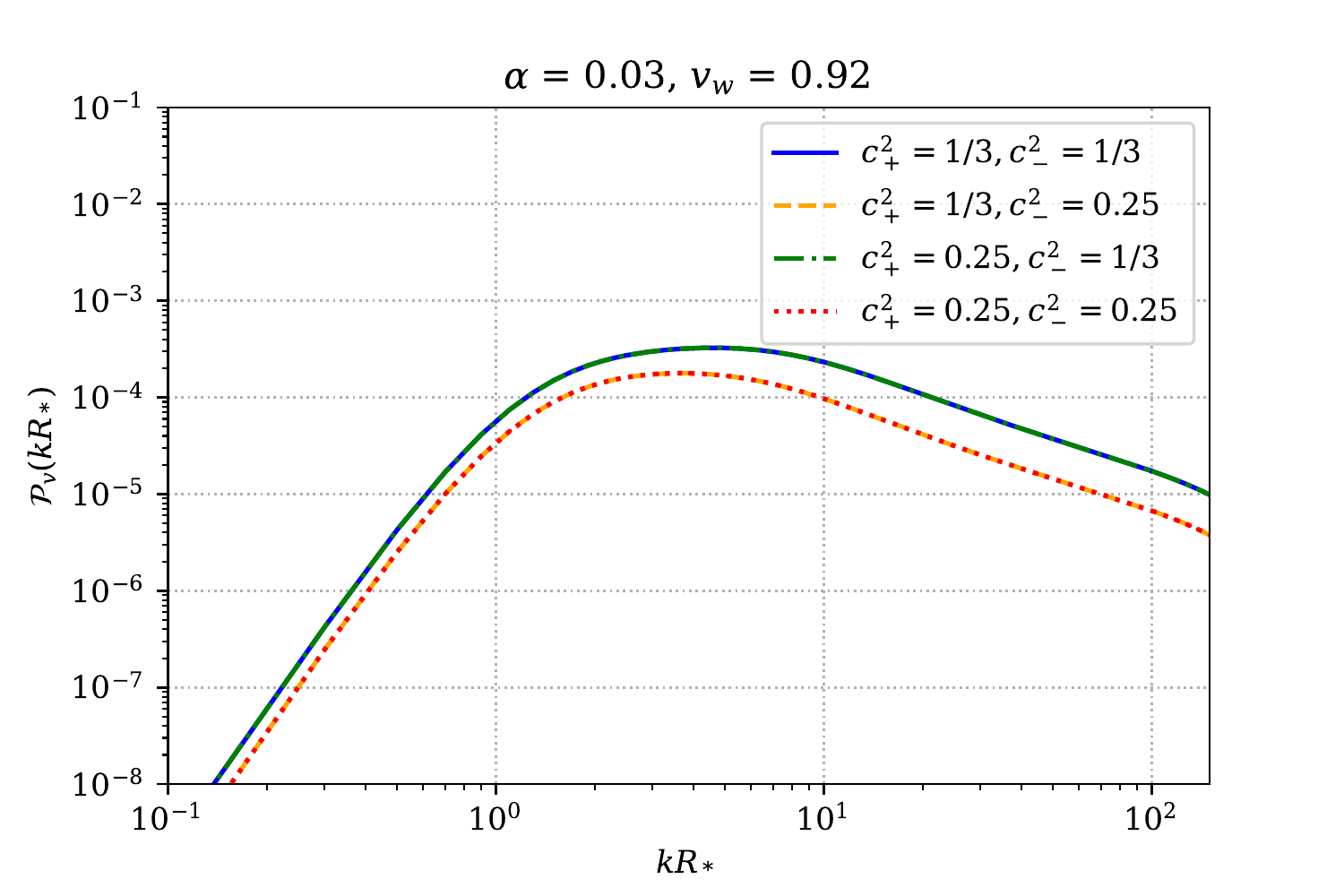}%
	\includegraphics[width=0.5\textwidth]{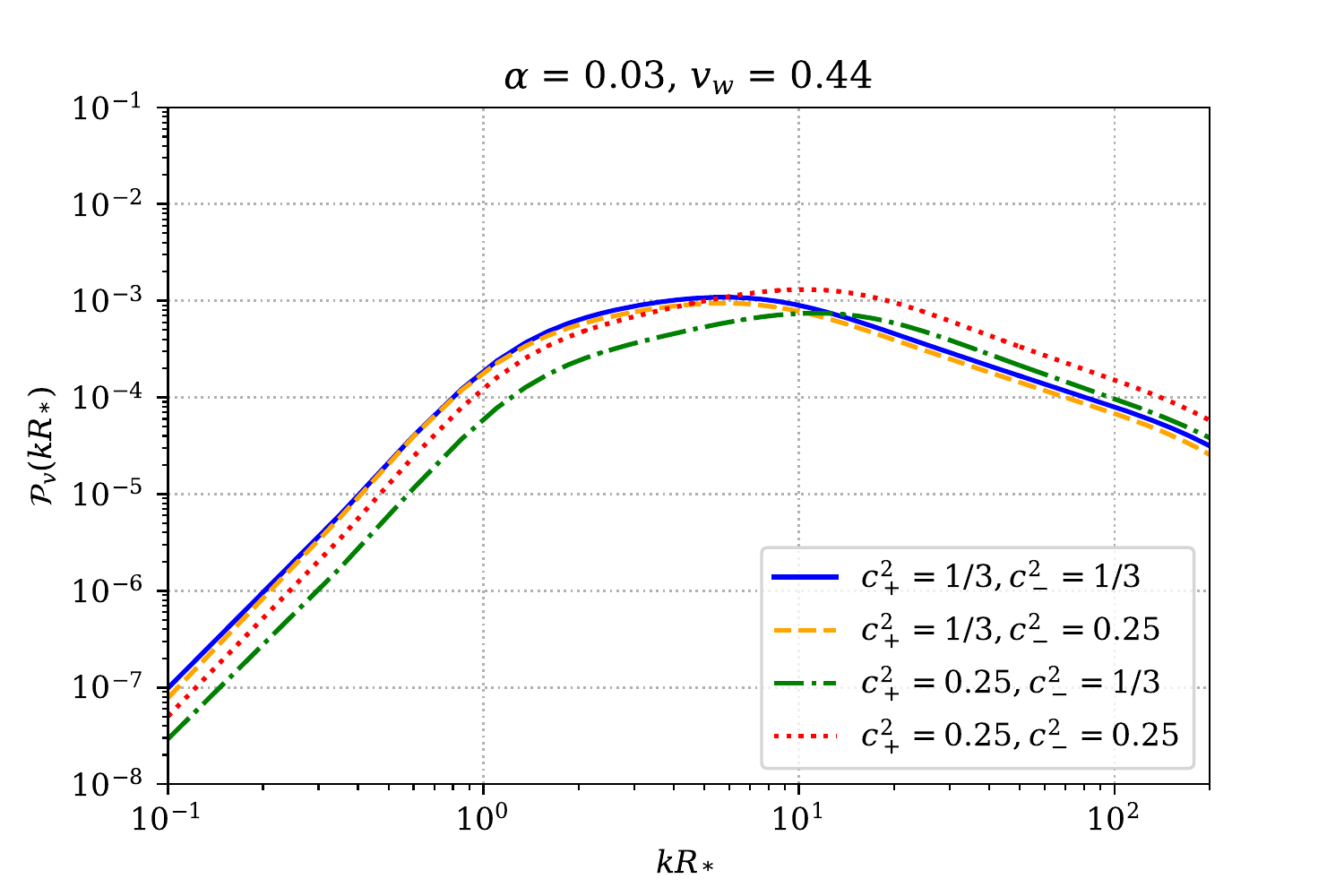}
	
	\subfigure[~Detonation]{
		\begin{minipage}[t]{0.5\textwidth}		\includegraphics[width=1\textwidth]{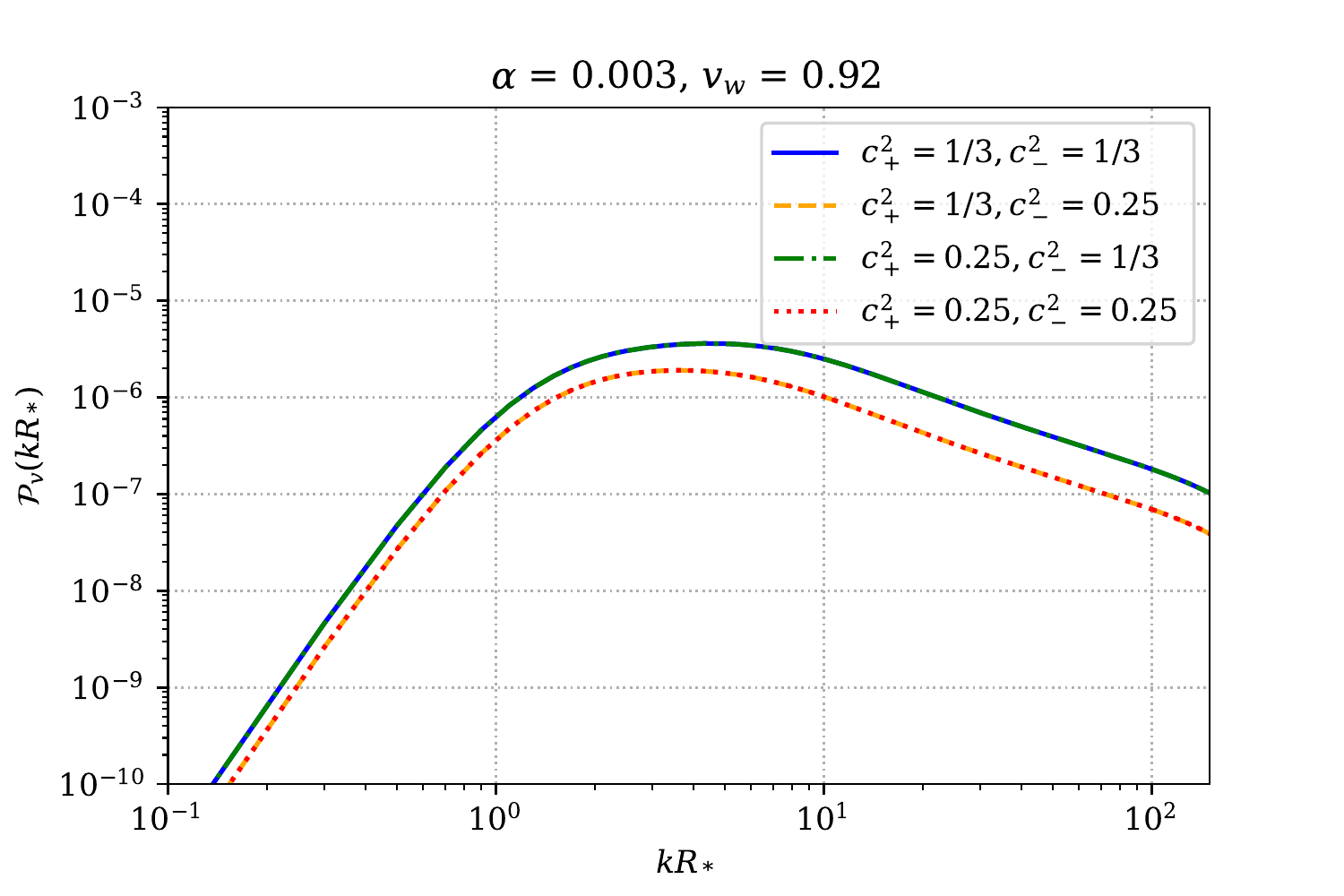}
			
	\end{minipage}}%
	\subfigure[~Deflagration]{
		\begin{minipage}[t]{0.5\textwidth}
			\includegraphics[width=1\textwidth]{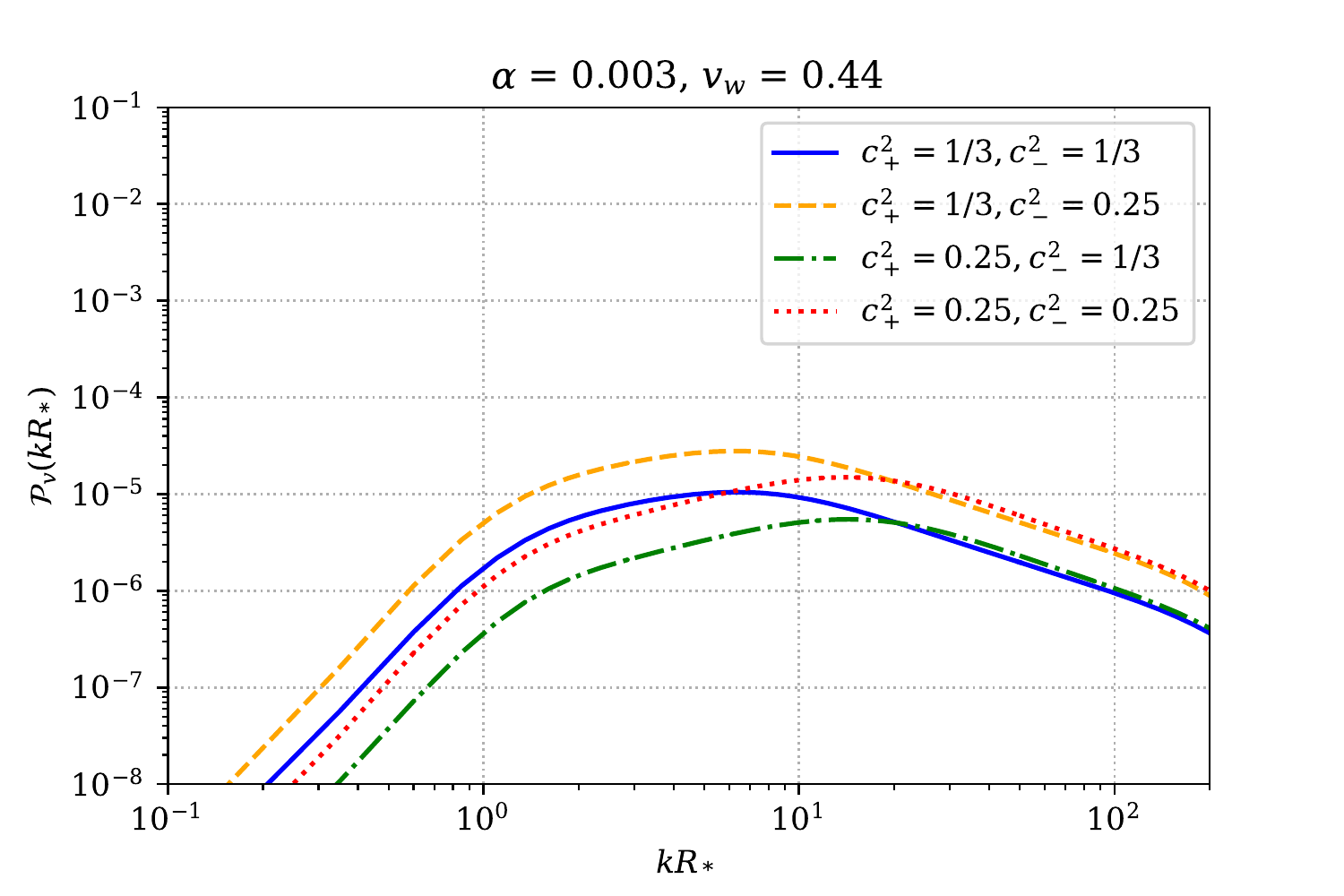}
	\end{minipage}}
	\caption{Sound velocity effects on velocity power spectra for detonation (left column) and deflagration(right column).}
	\label{spe_deto}
\end{figure*}

For $N$ bubbles in a given volume $\mathcal{V}$, the plane wave amplitude correlation function is 
\begin{equation}
\big\langle v_{\bf q}^iv_{\bf q'}^j\big\rangle = \sum_{m=1}^{N}\sum_{n=1}^{N}\bigg\langle\mathcal{T}_m^3\mathcal{T}_n^3\hat{z}^i\hat{z}^{\prime j}A(z)A(z')e^{-i\mathbf{q}\cdot\mathbf{x}_m + i\mathbf{q}'\cdot\mathbf{x}_n}e^{i(\omega_1 - \omega_2)t_i}\bigg\rangle\,\,.
\end{equation}
Averaging over the bubble nucleation sites in $[t',t' + dt']$, and the colliding time interval $[t_i, t_i + dt_i]$, we have
\begin{equation}
\sum_{m=1}^{N}\sum_{n=1}^{N}\bigg\langle e^{-i\mathbf{q}\cdot\mathbf{x}_m + i\mathbf{q}'\cdot\mathbf{x}_n}\bigg\rangle = d^2P\frac{N}{\mathcal{V}}(2\pi)^3\delta(\bf q - q')\,\,.
\end{equation}
Here $n(\mathcal{T}_i) = \frac{N}{\mathcal{V}}\frac{dP(\mathcal{T}_i)}{d\mathcal{T}_i}$ is the probability density distribution of bubble lifetime as mentioned in Sec.~\ref{SSM:PTdynamics}; then,
\begin{equation}
\big\langle v_{\bf q}^iv_{\bf q'}^j\big\rangle = \int d\mathcal{T}_in(\mathcal{T}_i)\mathcal{T}_i^6\hat{z}^i\hat{z}^j|A(z)|^2(2\pi)^3\delta(\bf q - q')\,\,.
\label{eq:vcorr}
\end{equation}
Substituting Eqs.~\eqref{eq:coltd1} and~\eqref{eq:coltd2} into Eq.~\eqref{eq:vcorr},
we can derive the spectral density of the velocity field as
\begin{equation}
P_v(q) = \frac{1}{\beta^6R_*^3}\int d\tilde{\mathcal{T}}f_{\rm col}(\tilde{\mathcal{T}})\tilde{\mathcal{T}}^6 \Bigg|A\left(\frac{\tilde{\mathcal{T}}q}{\beta}\right)\Bigg|^2\,\,,
\end{equation}
where $\tilde{\mathcal{T}} = \beta \mathcal{T}_i$.
Then, we can obtain the velocity power spectrum
\begin{equation}
\begin{split}
\mathcal{P}_v(q) &= 2\frac{q^3}{2\pi^2}P_v(q) \\
&= \frac{2}{(\beta R_*)^3}\frac{1}{2\pi^2}\left(\frac{q}{\beta}\right)^3\int d\tilde{\mathcal{T}} f_{\rm col}(\tilde{\mathcal{T}})\tilde{\mathcal{T}}^6\Bigg|A\left(\frac{\tilde{\mathcal{T}}q}{\beta}\right)\Bigg|^2\,\,.
\end{split}
\end{equation}

In Fig.~\ref{spe_deto}, we demonstrate the results of velocity power spectra for detonation (left column) and deflagration (right column) with different phase transition strengths. The effects of sound velocities on the velocity power spectra are obvious for both detonation and deflagration.
For detonation cases, the velocity power spectra can only be affected by the sound velocity of broken phase $c_-$, since the corresponding velocity and energy fluctuation profiles are only modified by $c_-$.
However, for deflagration, the velocity power spectra can be affected by both $c_-$ and $c_+$ and show more complicated behaviors.\footnote{More detailed studies are left for our future work.}

\section{Gravitational wave spectra for different initial conditions}\label{SSM:GWspectrum}
Having obtained the velocity profiles and hence the velocity power spectra, we could get
the velocity UETC and then the shear stress UETC.
Equation~\eqref{suetc} defines the shear stress UETC, which is determined by the source tensor $\tau^{ij}$.
The source tensor is dominated by the fluid velocity and it is $\tau_{ij} \simeq \bar{w}v_iv_j$ in the nonrelativistic situation.
Thus, in Fourier space 
\begin{equation}
	\tau_{i j}(\mathbf{k}, t)=\bar{w} \int d^{3} q \tilde{v}^{i}(\mathbf{q}, t) \tilde{v}^{j}(\tilde{\mathbf{q}}, t), \quad \tilde{\mathbf{q}}=\mathbf{q}-\mathbf{k}\,\,.
\end{equation}
Then we can schematically write the shear stress UETC in terms of  velocity UETC for the Gaussian velocity field as
\begin{tcolorbox}
\begin{equation}
P_{\dot{h}} \sim U_{\Pi} \sim	\langle\tau \tau\rangle \sim\langle \tilde{v} \tilde{v} \tilde{v} \tilde{v}\rangle=\sum \langle \tilde{v} \tilde{v} \rangle\langle \tilde{v}\tilde{v}\rangle  \sim \sum P_v P_v\,\,.
\end{equation}
\end{tcolorbox}

\begin{figure*}[!t]
	\centering
	\includegraphics[width=0.5\textwidth]{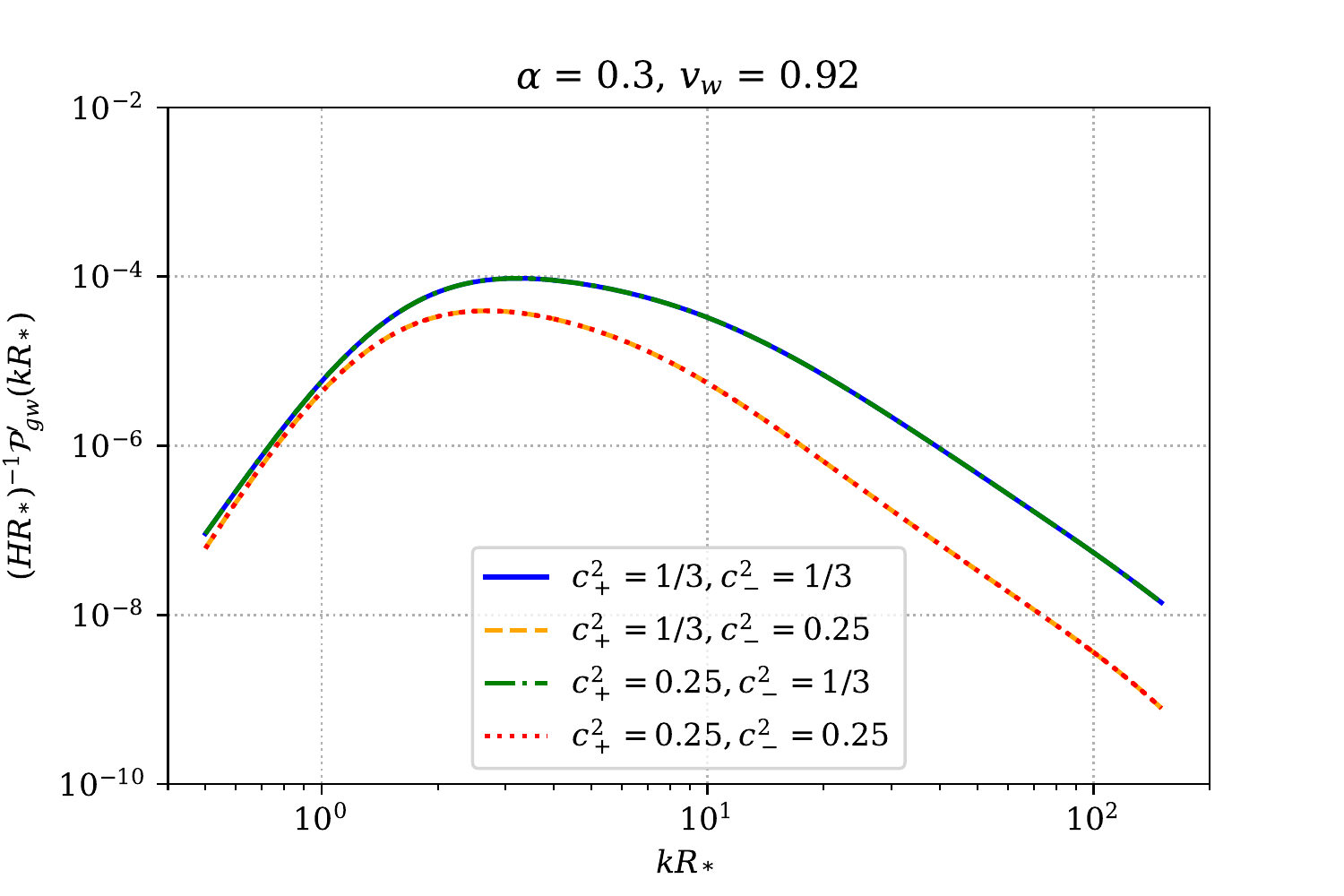}%
	\includegraphics[width=0.5\textwidth]{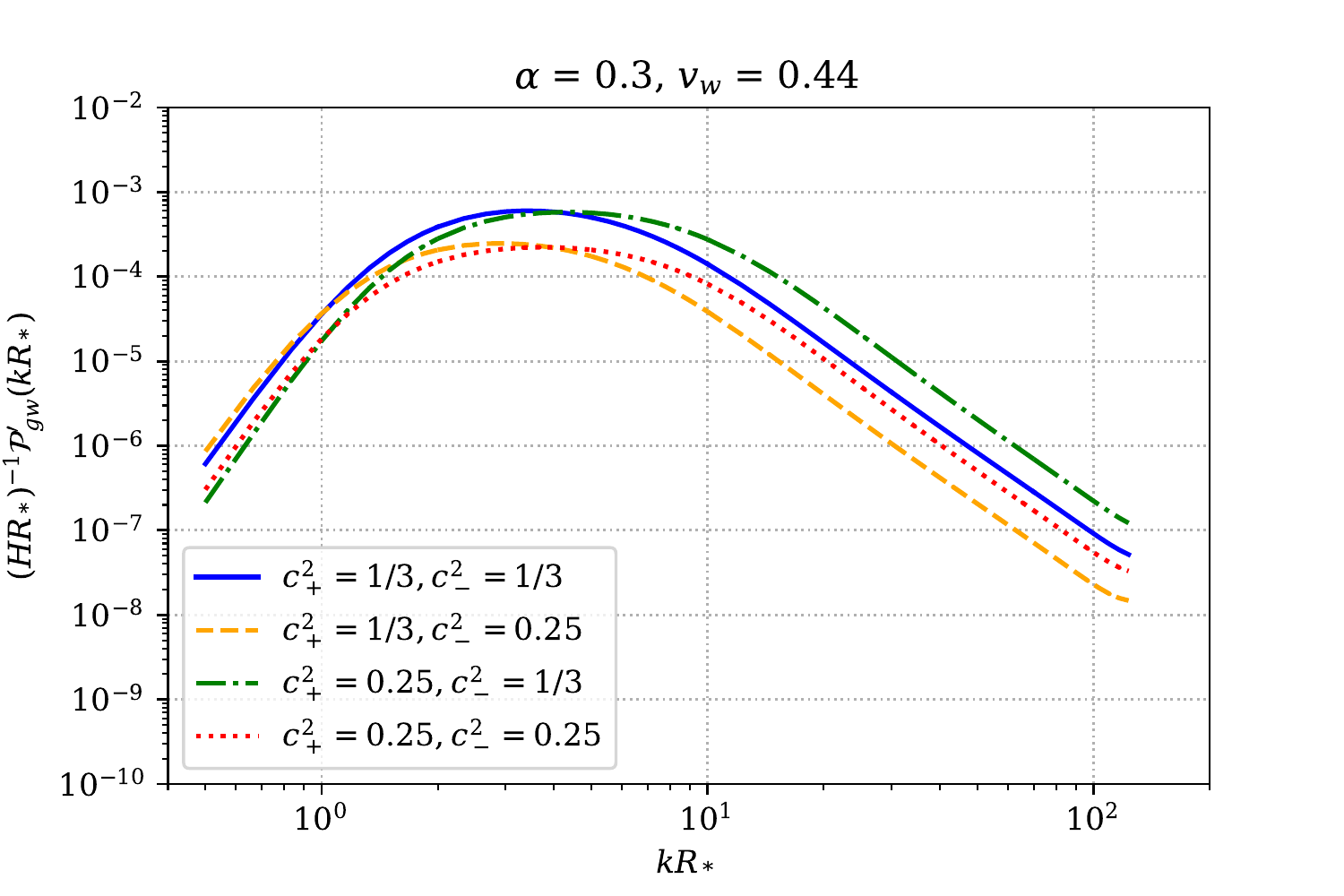}
	
	\includegraphics[width=0.5\textwidth]{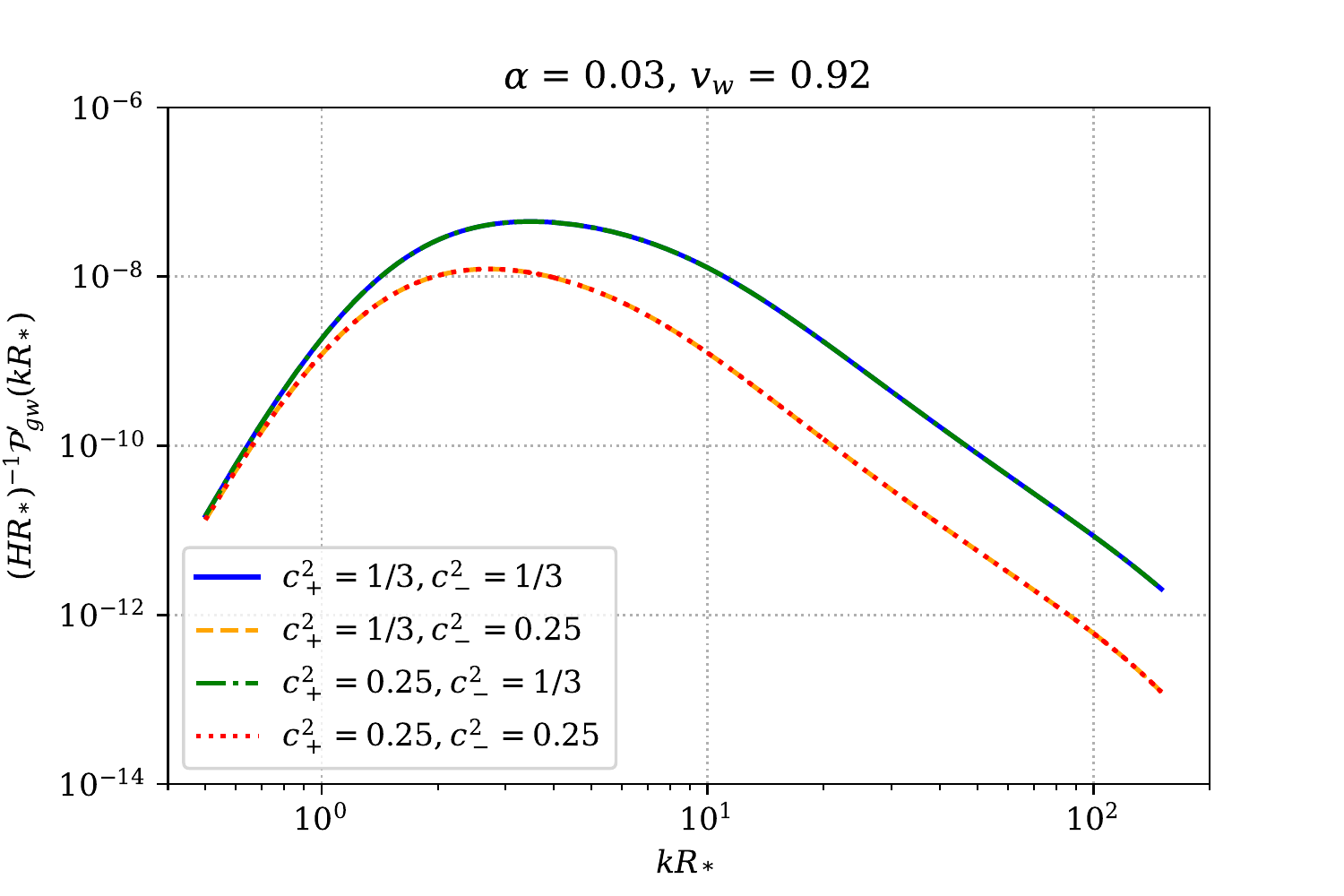}%
	\includegraphics[width=0.5\textwidth]{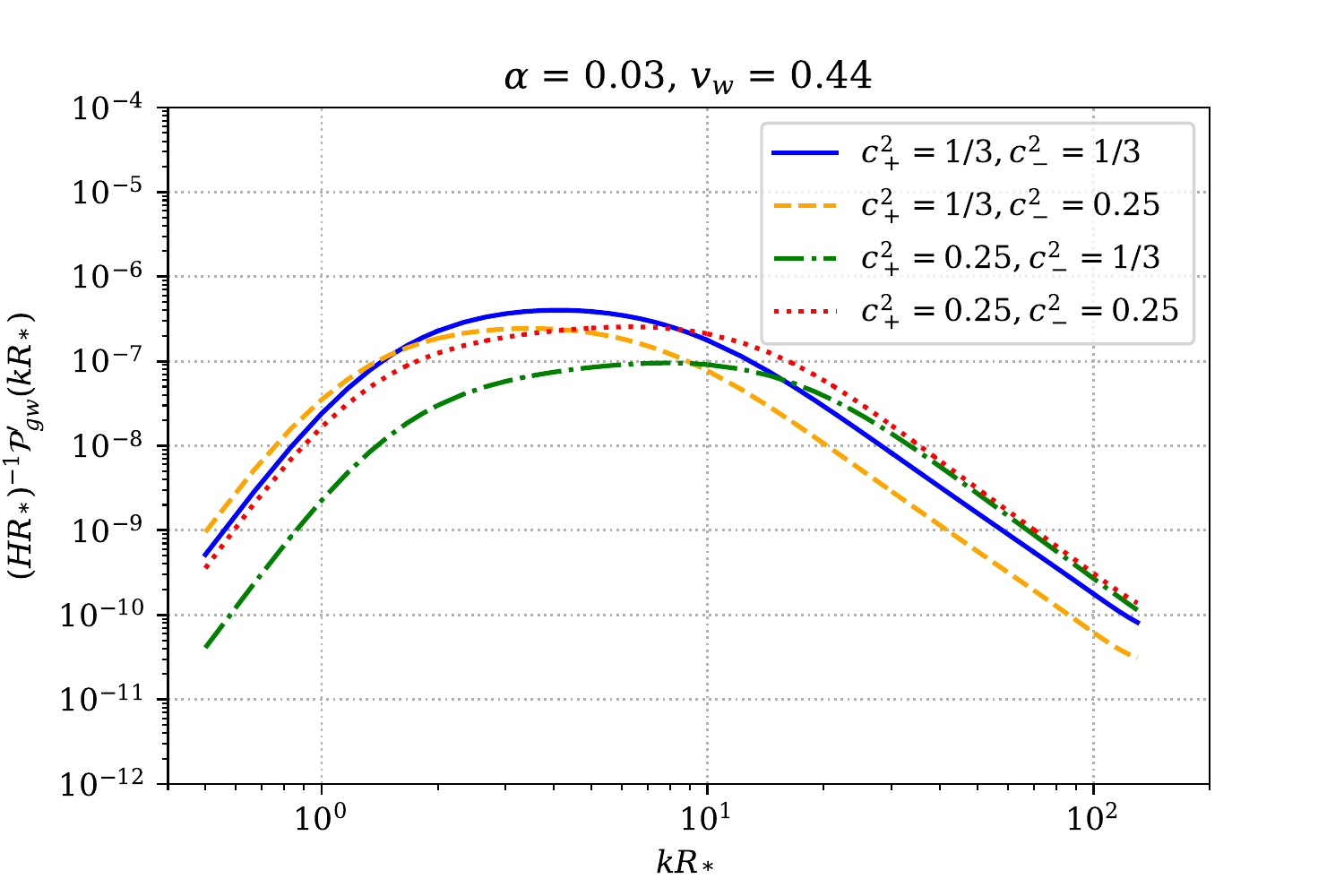}
	\subfigure[~Detonation]{
		\begin{minipage}[t]{0.5\textwidth}		\includegraphics[width=1\textwidth]{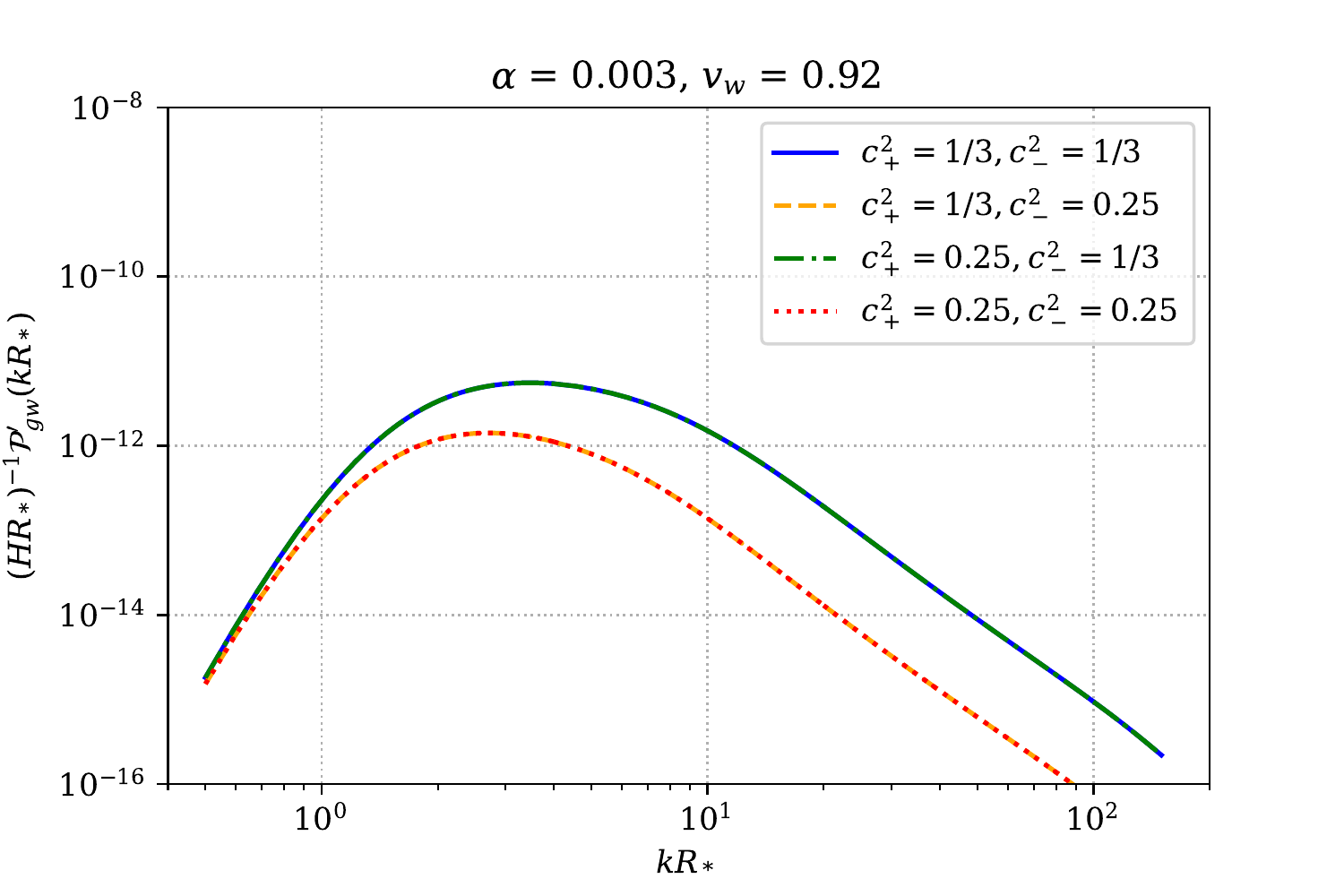}
	\end{minipage}}%
	\subfigure[~Deflagration]{
		\begin{minipage}[t]{0.5\textwidth}
			\includegraphics[width=1\textwidth]{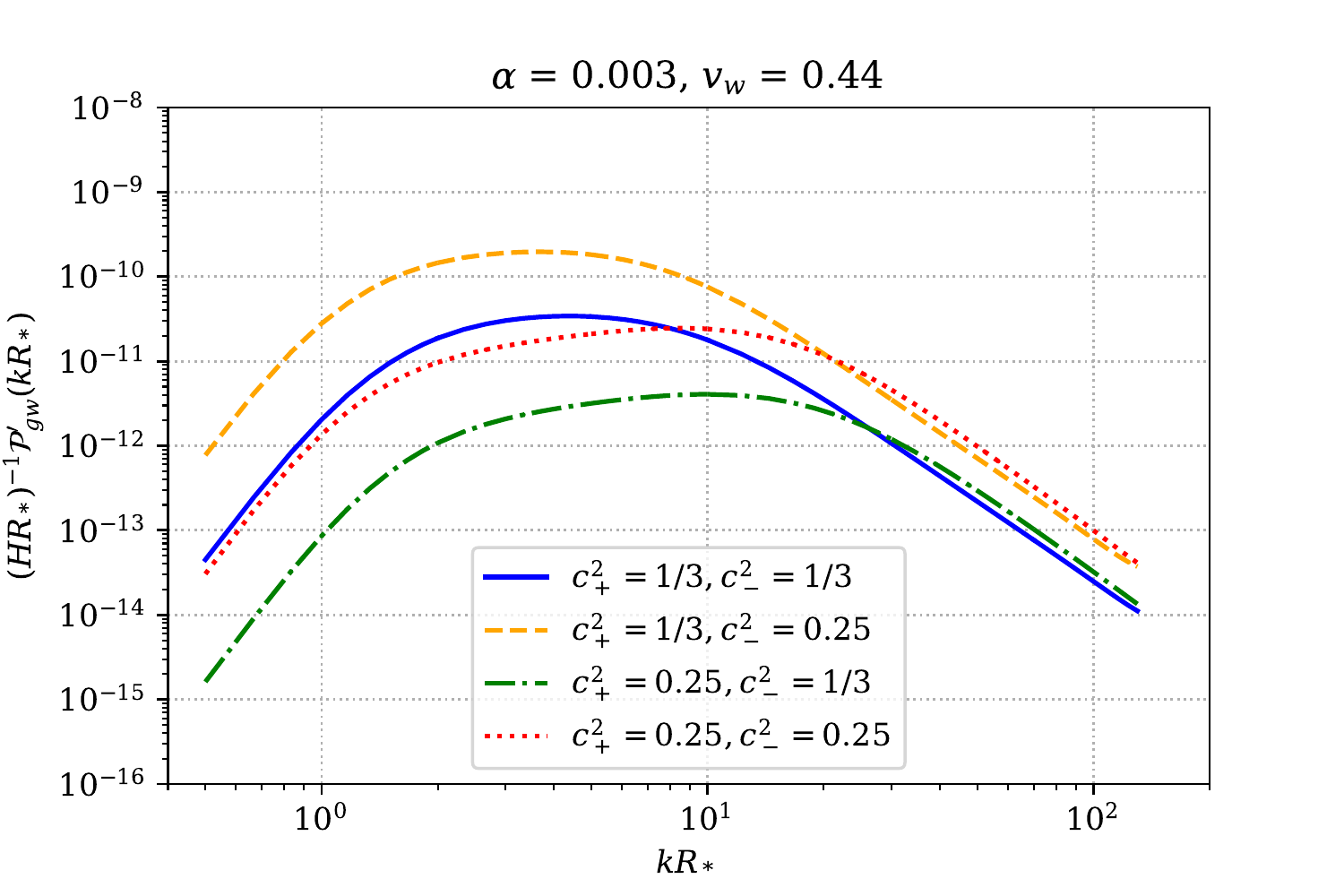}
	\end{minipage}}
	\caption{Sound velocity effects on the scaled GW power spectra for detonation (left column) and deflagration (right column).}\label{gwps}
\end{figure*}

After deriving $\langle \tilde{v}\tilde{v}\rangle$, the shear stress UETC can be obtained as 
\begin{equation}
U_{\Pi}(k,t_1,t_2) = 4\bar{w}^2\int\frac{d^3q}{(2\pi)^3}\frac{q^2}{\tilde{q}^2}(1 - \mu^2)^2P_v(q)P_v(\tilde{q})\cos(\omega t_-)\cos(\tilde{w}t_-)\,\,,
\end{equation}
with $t_- = t_1 - t_2$ and $\mu = \hat{\bf q}\cdot\hat{\bf k}$. For details, see Ref.~\cite{Hindmarsh:2019phv}.
Substituting the shear stress UETC into the  GW spectral  density formula in Eq.~\eqref{pmd}, we obtain the  dimensionless GW 
spectral density,
\begin{equation}
\begin{split}
\tilde{P}_{\rm gw}(y) =& \frac{1}{4\pi y c_s}\left(\frac{1 - c_s^2}{c_s^2}\right)\\
&\times\int_{z_-}^{z_+}  \frac{d z}{z}  \frac{(z - z_+)^2(z - z_-)^2}{z_+ + z_- - z}\bar{P}_v(z)\bar{P}_v(z_+ + z_- - z)\,\,,
\end{split}
\end{equation}
with $P_v(q) = L_f^3\overline{U}_f^2\bar{P}_v(qL_f)$. And $L_{\rm f}$ is the length scale of the velocity field.
Then using Eq.~\eqref{pgl}, we could obtain 
the GW power spectrum 
\begin{equation}
\mathcal{P}_{\rm gw}(k) = 3\left(\tilde{\Gamma}\overline{U}_{\rm f}^2\right)^2(H\tau_v)(HL_{\rm f})\frac{(kL_{\rm f})^3}{2\pi^2}\tilde{P}_{\rm gw}(kL_f)\,\,,
\label{Eq::gwproduc}
\end{equation}
where $y = kL_{\rm f}$, $z = qL_{\rm f}$, $z_\pm = y(1\pm c_s)/2c_s$, $\tau_v$ is the lifetime of sound wave, and the adiabatic index $\tilde{\Gamma} \approx 1 + c_s^2$.
The root-mean-square fluid velocity is
\begin{equation}
\begin{split}
\overline{U}_{\rm f}^2 &= \int\frac{dq}{q}\mathcal{P}_{\tilde{v}}(q)\\
&= \frac{2}{(\beta R_*)^3}\int d\tilde{\mathcal{T}}f_{\rm col}(\tilde{\mathcal{T}})\tilde{\mathcal{T}}^3\int dz\frac{z^2}{2\pi^2}|A(z)|^2\\ &= \frac{3}{4\pi v_w^3}\int dz\frac{z^2}{2\pi^2}2|A(z)|^2\,\,.
\end{split}
\end{equation}
We can also derive the growth rate of GW power spectrum scaled by the Hubble rate as
\begin{equation}
\mathcal{P}_{\rm gw}' = \frac{1}{H}\frac{d}{dt}\mathcal{P}_{\rm gw} = 3\left(\tilde{\Gamma}\overline{U}_{\rm f}^2\right)^2(HL_{\rm f})\frac{(kL_{\rm f})^3}{2\pi^2}\tilde{P}_{\rm gw}(kL_f)\,\,.
\end{equation}
When the GW was generated, the sound wave should freely propagate in the broken phase,
Hence the sound velocity $c_s$ in the calculation of GW should be $c_-$ in this work.
We can see that the sound velocity effects should be considered to predict more reliable GW signals.

\section{Discussions}\label{SSM:dis}

\begin{figure*}
	\centering
	\includegraphics[width=0.5\textwidth]{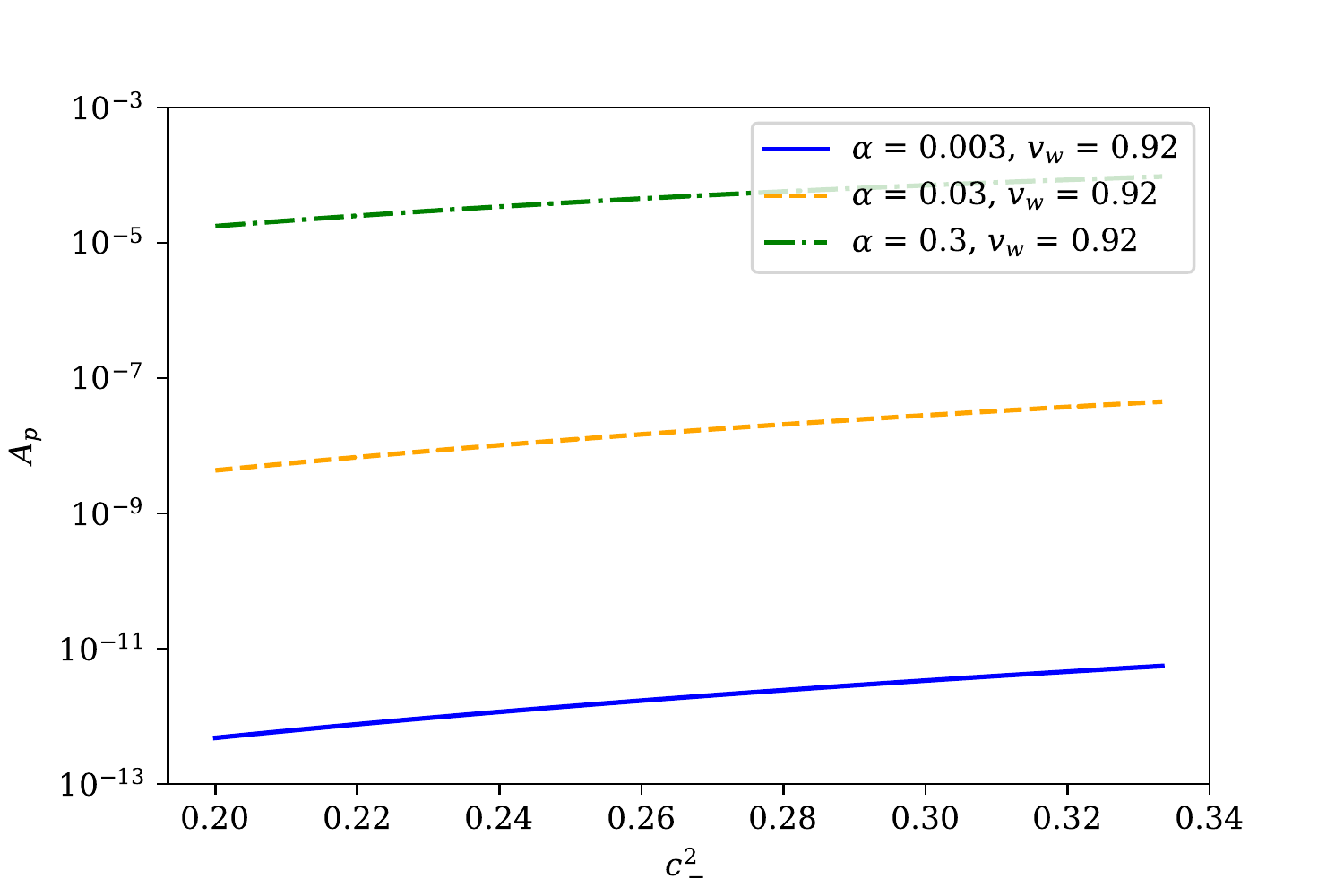}%
	\includegraphics[width=0.5\textwidth]{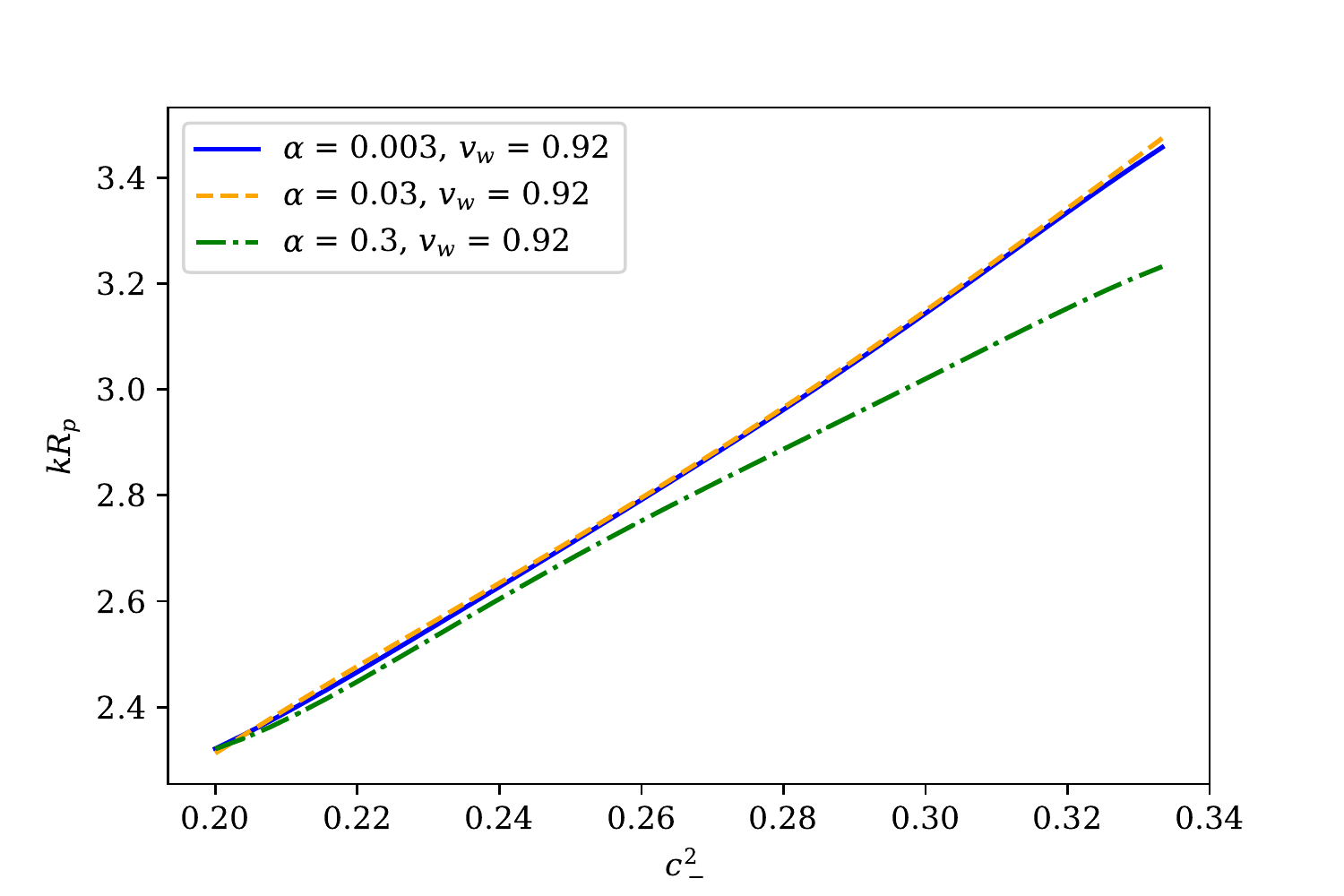}
	\caption{Peak amplitude (left panel) and peak angular frequency (right panel) of scaled GW spectrum as a function of sound velocities for detonation.}
	\label{Fig:dtkRAp}
\end{figure*}

\begin{figure*}
\centering
\includegraphics[width=0.5\textwidth]{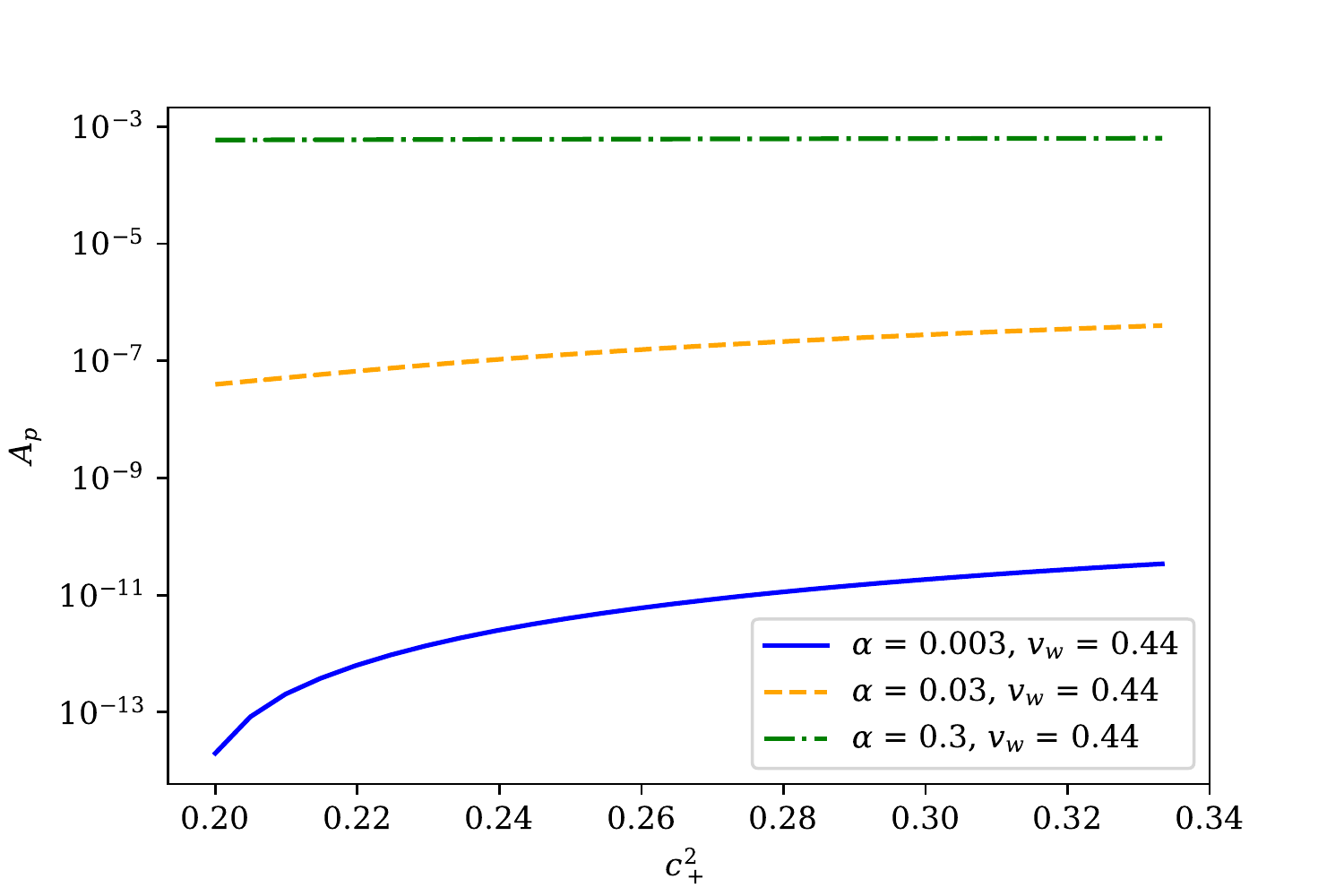}%
\includegraphics[width=0.5\textwidth]{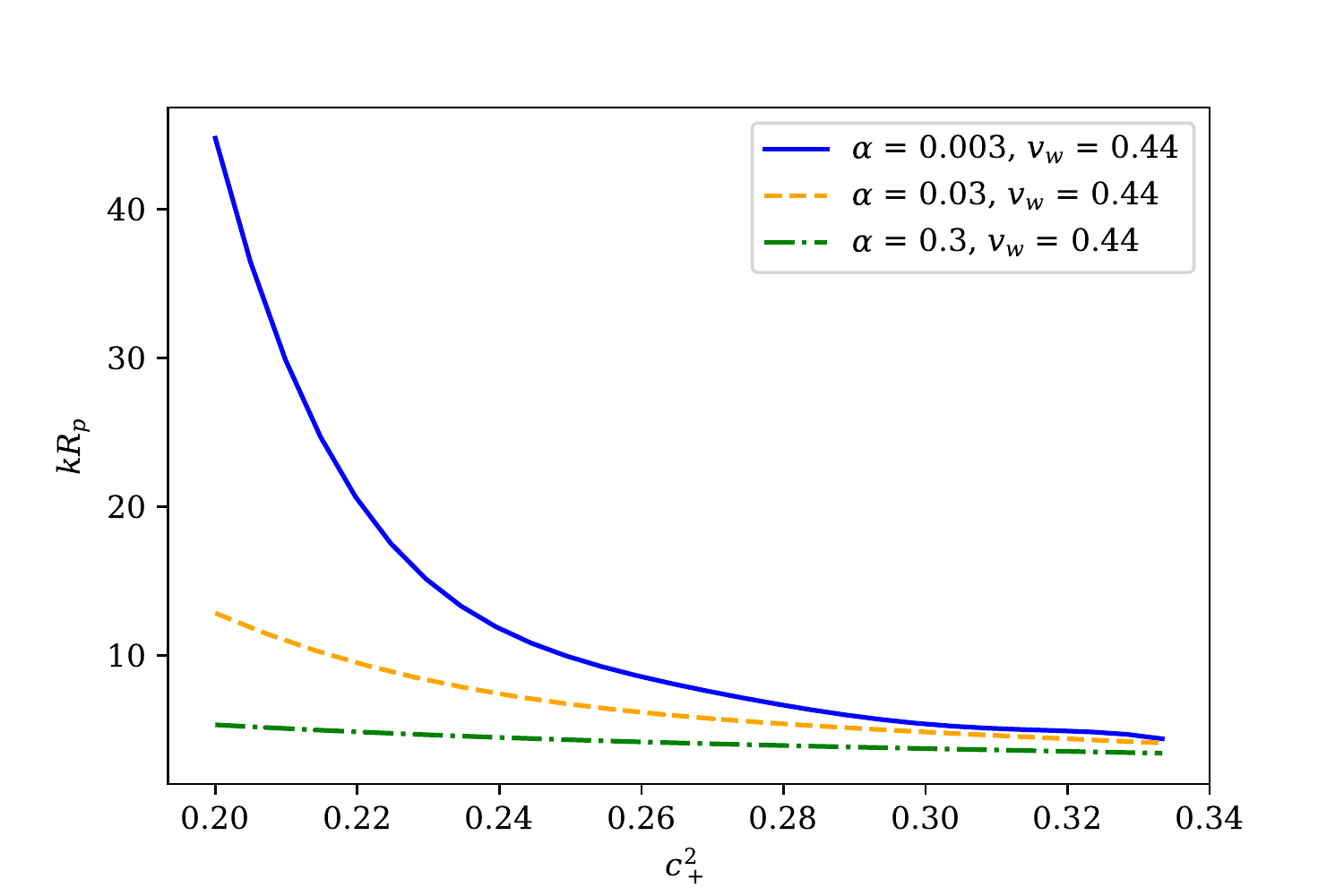}

\includegraphics[width=0.5\textwidth]{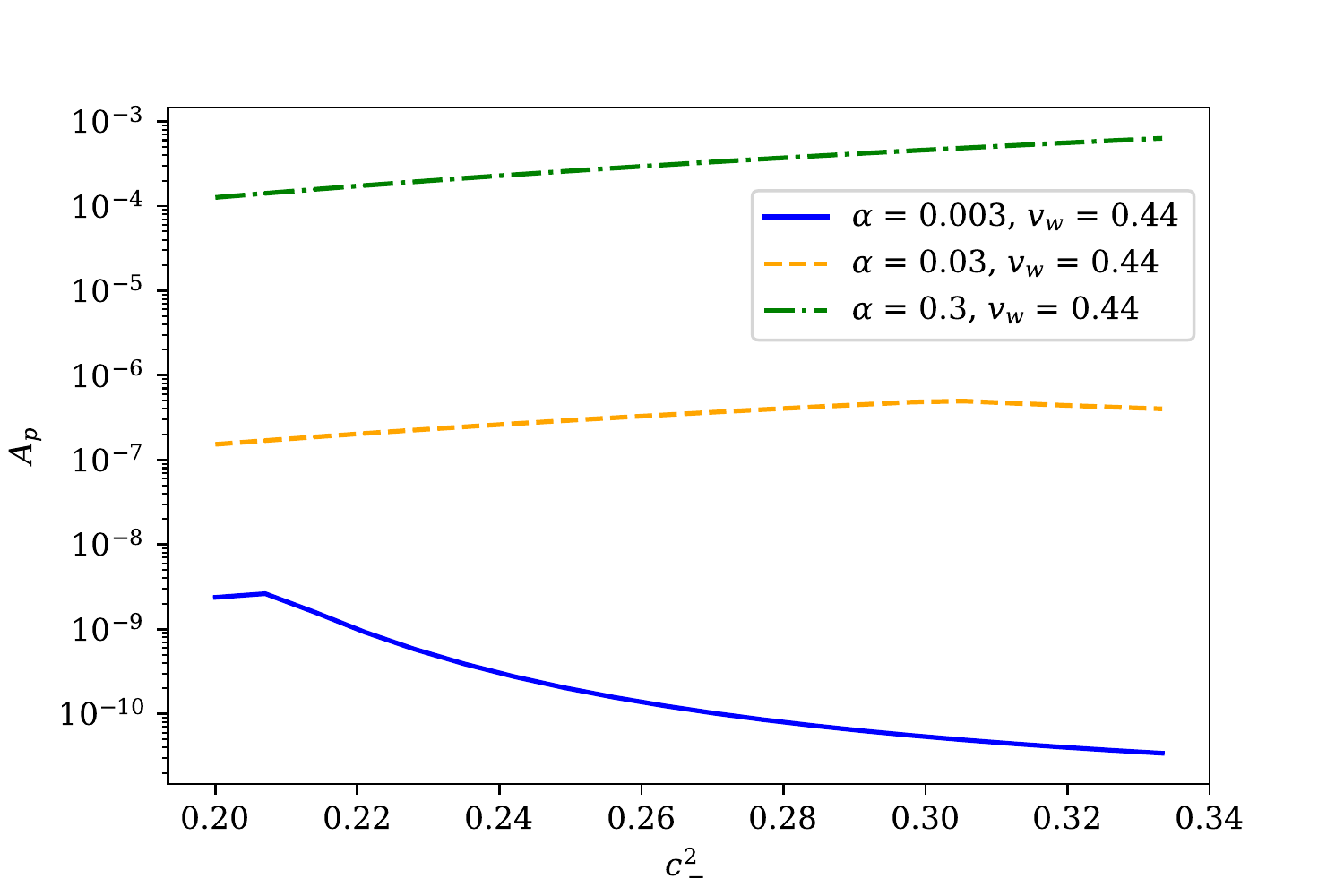}%
\includegraphics[width=0.5\textwidth]{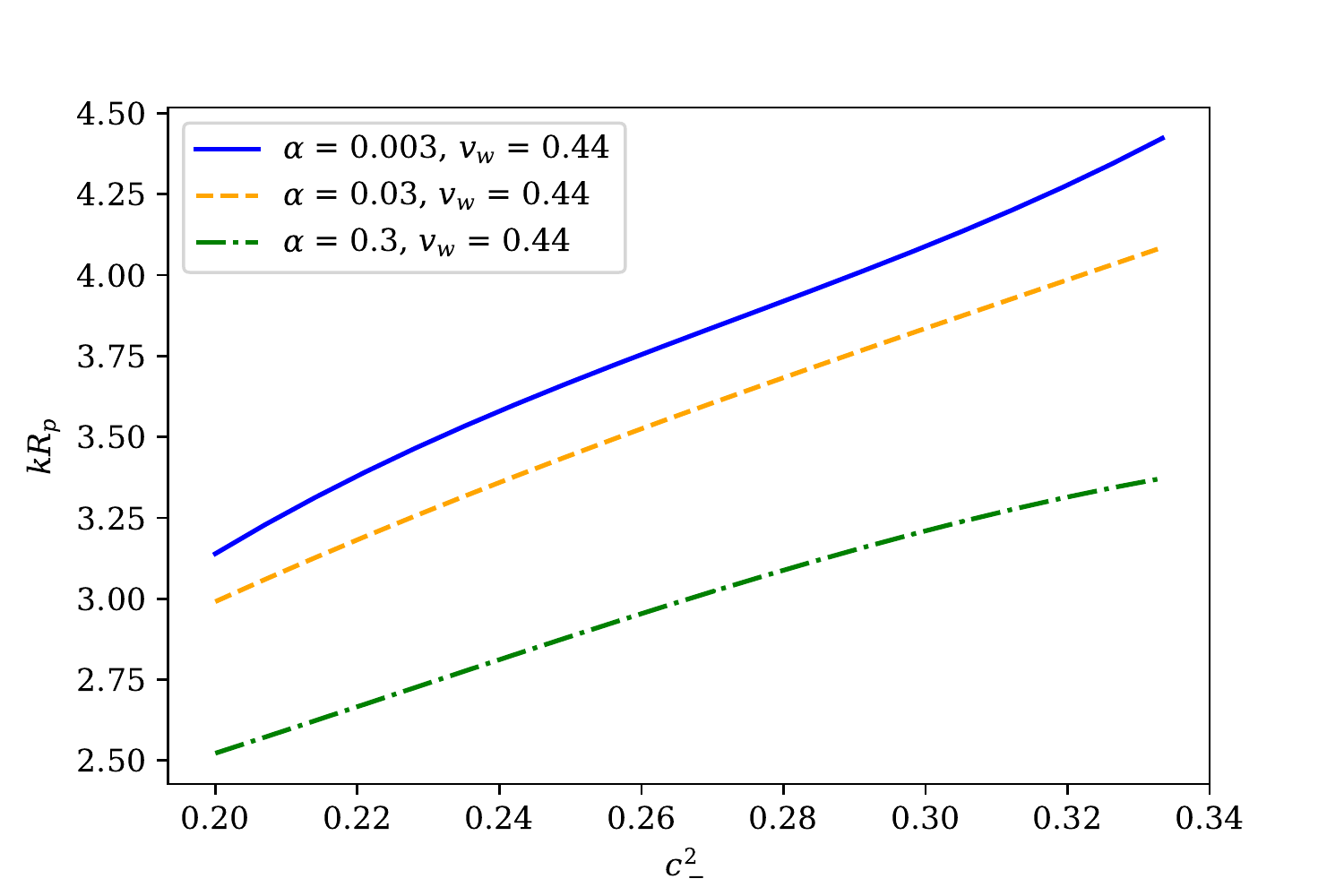}

\caption{Peak amplitude (left column) and peak angular frequency (right column) as a function of sound velocities for deflagration.}
\label{Fig:dfkRAp}
\end{figure*}

Using the detailed procedures mentioned above in the SSM, we present main results in Figs.~\ref{gwps}, \ref{Fig:dtkRAp}, and \ref{Fig:dfkRAp}.
In Fig.~\ref{gwps},  we show the scaled GW power spectra for detonation (left column) and deflagration (right column) with different phase transition strengths.
We can see that the peak amplitude and peak frequency of the scaled GW power spectra are modified after considering different sound velocities in symmetric and broken phases.
For detonation, only the sound velocity $c_-$ of the broken phase can give an obvious modification to the GW power spectra, since only $c_-$ can change the corresponding velocity and energy fluctuation profiles.
In contrast, the sound velocities of both symmetric and broken phases could affect the GW power spectra for deflagration cases.
The stronger dependence of the GW power spectrum on sound velocities for deflagration is strongly related to the corresponding fluid profiles derived by solving the hydrodynamical equations.
In our study, when the strength of phase transition is weaker, we actually found the fluid profile could get stronger dependence on sound velocities for deflagration.
Hence, the GW power spectrum derived by these fluid profiles shows stronger dependence on sound velocities, as shown in the bottom of Fig.~\ref{gwps}(b).
Recent numerical study~\cite{Cutting:2019zws} shows that there are kinetic defects as the phase transition strength increases in deflagration cases.
To take this effect into account in the SSM, Ref.~\cite{Gowling:2021gcy} introduces a suppression factor.
This factor can quantitatively change our result, but not qualitatively. We do not consider this suppression factor in this work.

In Fig.~\ref{Fig:dtkRAp}, we show the sound velocity effects on the peak amplitude $A_p$ and the peak angular frequency $kR_p$ of the scaled GW spectrum for detonation.
Here, we use different line styles with different colors to represent the results obtained by different phase transition strengths.
According to both panels of Fig.~\ref{Fig:dtkRAp}, we find larger sound velocity differences produce larger deviation of the peak amplitude and the peak angular frequency.  
With the decreasing of the sound velocity of broken phase, the peak amplitude and the peak angular frequency monotonically decrease.
Since the sound velocity of symmetric phase could not affect the GW power spectra of detonation, we set $c_+ = 1/\sqrt{3}$ here.

However, for deflagration, the behaviors of peak amplitude and peak angular frequency modification are relatively complicated.
We show the peak amplitude and peak angular frequency as function of sound velocities for deflagration in Fig.~\ref{Fig:dfkRAp}. Different lines with different colors denote different phase transition strengths.
In the top panels, we set $c_- = 1/\sqrt{3}$ and show effects of different sound velocities of the symmetric phase on the peak amplitude and the peak angular frequency, respectively.
With the decreasing of the sound velocity of the symmetric phase, the peak amplitude of GW spectrum monotonically decreases, while the peak angular frequency monotonically increases. 
For a weaker phase transition, the effect of sound velocity is more significant.
However, fixing the sound velocity $c_+ = 1/\sqrt{3}$ in the symmetric phase, the peak amplitude shows a relatively more complicated behavior with the decreasing of sound velocity in the broken phase as depicted in the bottom left panel of Fig.~\ref{Fig:dfkRAp}. 
We find the peak amplitude first increases and then decreases when the sound velocity of the broken phase is smaller than an inflection point for $\alpha = 0.003$ and $\alpha = 0.03$.
But with the increasing of phase transition strength, this inflection point of sound velocity becomes larger.
And when phase transition is strong enough, the peak amplitude monotonically decreases with the decreasing of the sound velocity in the broken phase.
For the peak angular frequency, it monotonically decreases with the decreasing of the sound velocity in the broken phase for different phase transition strengths.
As shown in Figs.~\ref{spe_deto} and \ref{Fig:dfkRAp}, the GW spectrum of deflagration is strongly related to the sound velocities of both phases and phase transition strength.

For the GW power spectrum, these modifications of peak amplitude and peak angular frequency originate from the variation of velocity and energy fluctuation profiles after considering different sound velocities in the DSVM.
These results remind us that we should precisely model the EoS in both symmetric and broken phases and calculate the sound velocities to get more precise predictions on the velocity and energy fluctuation profiles. 
Then, we could get more reliable velocity power spectra and hence more precise acoustic GW spectra.

According to the results shown in Figs.~\ref{gwps}, \ref{Fig:dtkRAp}, and \ref{Fig:dfkRAp}, we  find that  the sound velocity slightly changes the shape of the GW spectrum but mostly affects the peak amplitude and peak frequency for both the weak and strong phase transition.
Sound velocity and bubble wall velocity are degenerate with other phase transition parameters: the phase transition strength $\alpha$, duration of the phase transition $\beta$, and the phase transition temperature $T$.
However, the degeneracy of these parameters is difficult to quantitatively model here. It deserves a further study to reveal the degeneracy between the sound velocity, bubble wall velocity $v_w$, and other phase transition parameters, e.g. $\alpha$, $\beta$, and $T$.
Based on the results shown in Figs.~\ref{gwps} and \ref{Fig:dtkRAp}, a change of the sound velocity $c_-^2$ from $1/3$ to $0.25$ at $\alpha=0.3$ can have an impact of around $60\%$ on the peak amplitude and around $15\%$ on the peak frequency for detonation.
And for weaker phase transition (e.g. $\alpha = 0.03$ and $\alpha = 0.003$), a change of the sound velocity of the detonation mode can have an impact almost the same fraction of corrections on the peak amplitude and peak frequency as the $\alpha = 0.3$ case.
For deflagration, Figs.~\ref{gwps} and \ref{Fig:dfkRAp} show that a change of the sound velocity $c_+^2$ from $1/3$ to $0.25$ at $\alpha = 0.3$ can have an impact of around few percent on the peak amplitude and around $40\%$ on the peak frequency 
and a change of the sound velocity $c_-^2$ from $1/3$ to $0.25$ at $\alpha = 0.3$ can have an impact of around $63\%$ on the peak amplitude and around $15\%$ on the peak frequency.
For weaker phase transition (e.g. $\alpha = 0.03$ and $\alpha = 0.003$), a change of velocity $c_+^2$ from $1/3$ to $0.25$ can have an stronger impact on the peak amplitude and peak frequency for deflagration.
And a change of velocity $c_-^2$ from $1/3$ to $0.25$ can have an impact almost the same fraction of corrections on peak frequency as the $\alpha = 0.03$ case.
However, the impact on the peak amplitude becomes stronger and more complicated for the change of $c_-^2$.
Nevertheless, the impact from the sound velocity at $\alpha\sim\mathcal{O}(0.1)$ is already quite obvious. 
Since LISA  and TianQin will be sensitive to phase transitions with $\alpha\sim\mathcal{O}(0.1)$, sound velocity might be an important factor for the future GW experiments.

\subsection{Observational consequences }

\begin{figure*}[!ht]
	\centering
	\subfigure[~Detonation]{
		\begin{minipage}[t]{0.5\textwidth}		\includegraphics[width=1\textwidth]{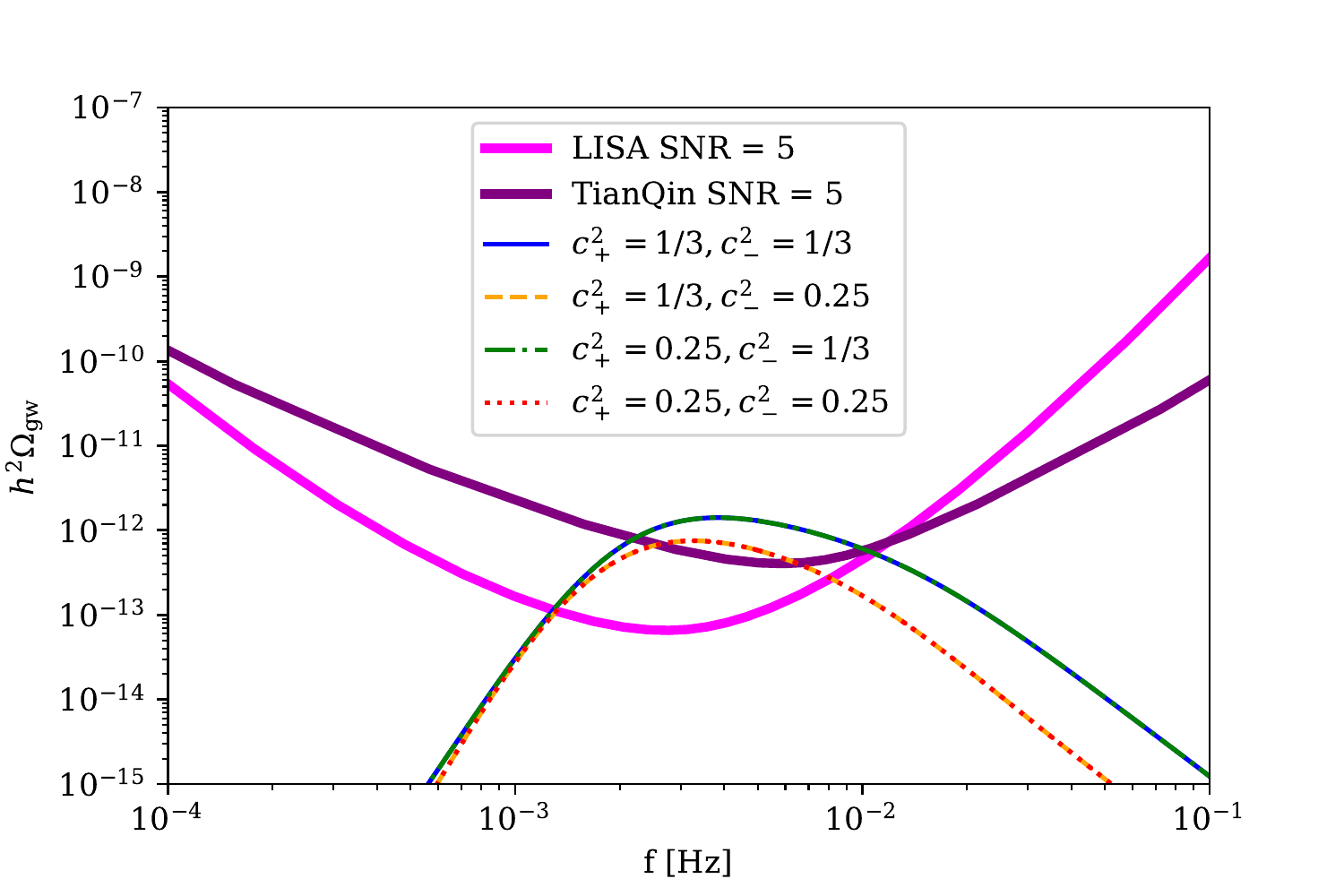}
	\end{minipage}}%
	\subfigure[~Deflagration]{
		\begin{minipage}[t]{0.5\textwidth}
			\includegraphics[width=1\textwidth]{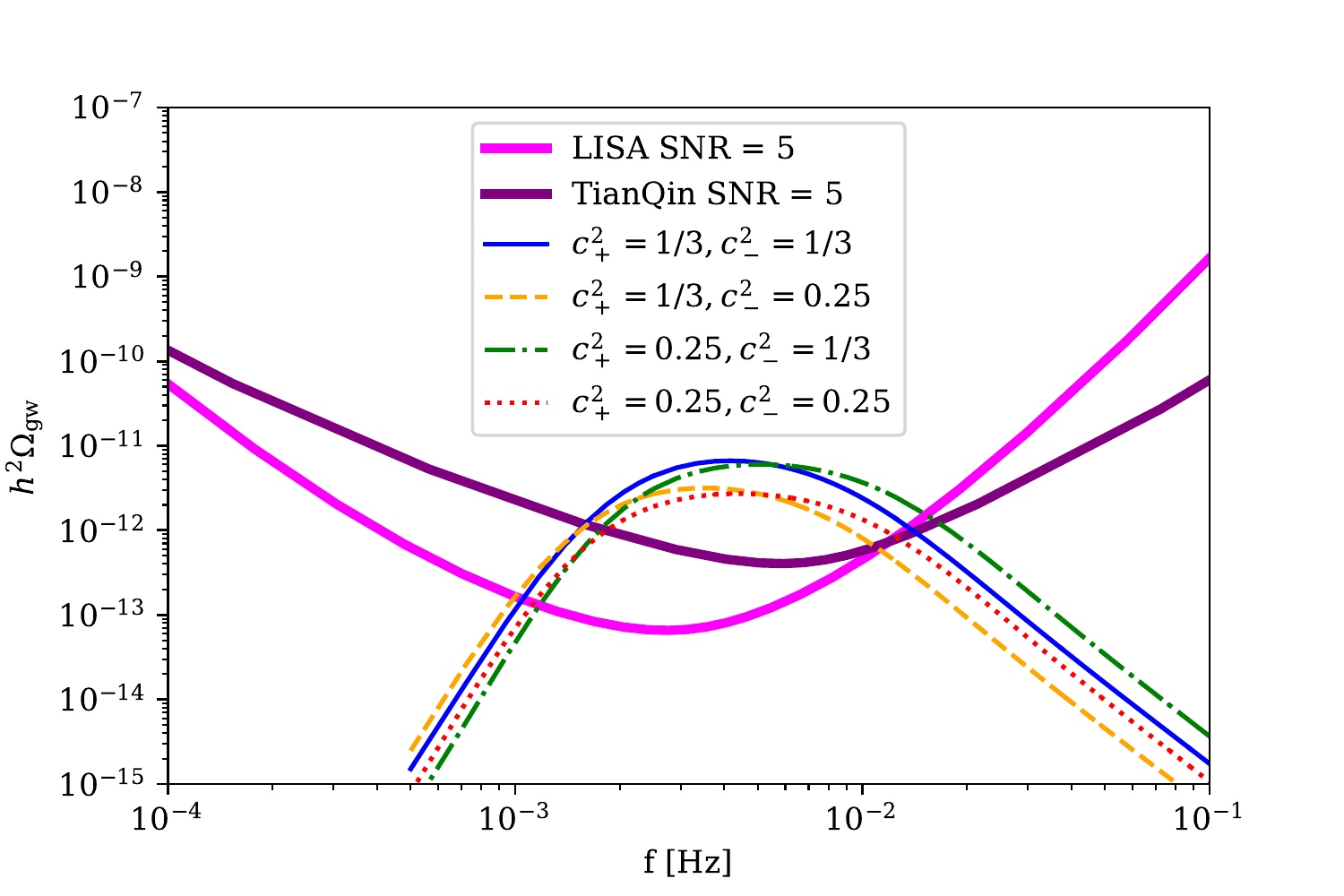}
	\end{minipage}}
	\caption{The  GW spectra for the benchmark set in Eq.~\eqref{Eq::bset} with different bubble wall velocities. The Left panel is for detonation with $v_w = 0.92$ and the right panel is for deflagration with $v_w = 0.44$. The solid ($c_+^2=1/3, c_-^2=1/3$), dashed ($c_+^2=1/3, c_-^2=0.25$), dot-dashed ($c_+^2=0.25, c_-^2=1/3$), and dotted ($c_+^2=0.25, c_-^2=0.25$) lines represent GW spectra obtained by different combinations of sound velocities. The magenta line and purple line are the power-law integrated sensitivities of LISA and TianQin at the level of  $\mathrm{SNR}=5$, respectively.}
	\label{Fig::GW0}
\end{figure*}

After we obtain the GW spectra at the production time,  it is straightforward to calculate the GW signals observed today and the
signal-to-noise ratio for a given GW experiment.  
We just need to calculate the concrete phase transition parameters and include the redshift effect.
And the sound velocity effects would directly reflect on the detectability of the PTGW signal.  
To clearly show  the effects of sound velocities on the GW detectability,
we demonstrate a simple example with the following benchmark set
\begin{equation}
	T_* = 500~\mathrm{GeV},\quad HR_* = 0.01,\quad \alpha = 0.4 \,\,.
	\label{Eq::bset}
\end{equation}
The GW power spectrum predicted by the SSM is a function of phase transition parameters $(T_*, \alpha, HR_*, v_w)$ and the scaled wave number $z = kR_*$ at the production time.
While the GW power spectrum, which is represented in terms of frequency $f$ today, can be derived as
\begin{equation}
	\Omega_{\mathrm{gw}}(f) = F_{\rm gw}\mathcal{P}_{\mathrm{gw}}(z(f))\,\,,
\end{equation}
where \cite{Hindmarsh:2017gnf}
\begin{equation}
	F_{\rm gw} = (3.57\pm0.05)\times 10^{-5}\left(\frac{100}{g_*}\right)^{1/3}\,\,.
\end{equation}
Taking the redshift into account, we have the following relation between $z$ and $f$
\begin{equation}
	f = \frac{z}{HR_*}f_{*}\,\,,
\end{equation}
where \cite{Caprini:2019egz}
\begin{equation}
	f_{*} = 2.6\times 10^{-6}~\mathrm{Hz}\left(\frac{T_*}{100~\rm GeV}\right)\left(\frac{g_*}{100}\right)^{1/6}\,\,.
\end{equation}
Finally, we have
\begin{equation}
	\Omega_{\mathrm{gw}}(f) = 3F_{\rm gw}\left(\tilde{\Gamma}^2\overline{U}_{\rm f}^3\right)(HR_*)^2\frac{(z)^3}{2\pi^2}\tilde{P}_{\rm gw}(z)\,\,.
\end{equation}
Here, we replace $H\tau_v$ in Eq.~\eqref{Eq::gwproduc} with $HR_*/\overline{U}_{\rm f}$~\cite{Caprini:2019egz,Hindmarsh:2017gnf}.
And we conventionally use $h^2\Omega_{\mathrm{gw}}$ ($h\approx0.678$) to represent the GW power spectrum in Fig.~\ref{Fig::GW0}.
The left and right panels of Fig.~\ref{Fig::GW0} denote the GW spectra of detonation (left panel $v_w=0.92$) and deflagration (right panel $v_w=0.44$) for the benchmark set in Eq.~\eqref{Eq::bset}. The solid ($c_+^2=1/3, c_-^2=1/3$), dashed ($c_+^2=1/3, c_-^2=0.25$), dot-dashed ($c_+^2=0.25, c_-^2=1/3$), and dotted ($c_+^2=0.25, c_-^2=0.25$) lines represent GW spectra derived by different combinations of sound velocities. 
The magenta line and purple line are the power-law integrated sensitivities of LISA and TianQin at the level of  $\mathrm{SNR}=5$ with $10^8~s$ observation time~\cite{Caprini:2019pxz,Liang:2021bde}, respectively.
As shown in Fig.~\ref{Fig::GW0}, different sound velocities can also affect the peak amplitude and peak frequency of PTGW observed today. Hence sound velocity could be important to the detectability of PTGW.
Therefore, more precise GW spectra depend on more reliable sound velocities of both phases.   
In Appendix~\ref{App:1}, we show the sound velocity could deviate from $1/\sqrt 3$ in a generic class of new physics models from the perspective of standard model effective field theory.

\section{conclusion}\label{SSM:con}
We have studied the sound velocity effects on the scaled gravitational wave spectra in the sound shell model.
We find that large deviation of sound velocities from the pure radiation value $1/\sqrt{3}$ leads to obvious shifts of peak amplitude and peak angular frequency.
To extract reliable information of the early Universe from the future gravitational wave experiments,
it is necessary to first obtain the precise sound velocities in symmetric and broken phases.
More comprehensive calculations in some representative new physics models using the above method are in progress.

\begin{acknowledgments}
X.W. and F.P.H. are funded by Guangdong Major Project of Basic and Applied Basic Research (Grant No. 2019B030302001).
F.P.H. is supported in part by SYSU startup funding.
\end{acknowledgments}

\appendix
\section{Sound velocity deviation in a representative  effective model}\label{App:1}

	We have shown that the sound velocities in the broken and symmetric phases could make significant modification on the peak amplitude and peak frequency of the PTGW.
	The sound velocity deviation from $1/\sqrt{3}$ of the pure radiation phase might appear in a wide classes of new physics models including the various examples in Ref.~\cite{Giese:2020rtr} and a representative effective model in Ref.~\cite{Wang:2020nzm}.
	To illustrate the common feature of sound velocity deviation in various new physics models, 
	we consider the representative effective model, namely, the dimension-6 effective model with the tree-level potential $V(\phi)=\frac{\mu^{2}}{2} \phi^{2}+\frac{\lambda}{4} \phi^{4}+\frac{\kappa}{8 \Lambda^{2}} \phi^{6}$. 
	From the perspective of standard model effective field theory,
	the effective model with dimension-6 operator could represent  various  new physics models, including the inert singlet, doublet, triplet, or composite Higgs model as discussed in our previous studies~\cite{Cao:2017oez,Huang:2015izx,Huang:2017rzf, Huang:2019riv,Wang:2020wrk}.
	Here, we take the approximated free energy of the effective model 
	\begin{equation}
	\mathcal{F}(\phi,T)=V_{\rm eff}(\phi,T)\approx - \frac{a_{\pm}}{3}T^4 + \frac{\mu^2 + cT^2}{2}\phi^2 + \frac{\lambda}{4}\phi^4 + \frac{\kappa}{8\Lambda^2}\phi^6  \,\,
	\end{equation}
	to demonstrate the sound velocity deviation.
	$a_\pm = g_\pm\pi^2/30$ ($g_\pm$ is the number of degrees of freedom),
	$\Lambda/\sqrt{\kappa}$ is the effective cutoff scale and $cT^2$ represents the thermal correction with
	\begin{equation}\label{cdim6}
	c=\frac{1}{16}(g^{\prime 2}+3 g^2+4 y_t^2+4 \frac{m_h^2}{v^2}-12 \frac{\kappa v^2}{\Lambda^2}) \,\,,
	\end{equation}
	where $g^{\prime}$ and $g$ are gauge couplings, $y_t$ is the Yukawa coupling of top quark, and $v$ is the electroweak vacuum expectation value.
	Hence, we have the pressure $p = - \mathcal{F}$ and energy density $e = T\frac{\partial p}{\partial T} - p$.
	Since the definition of sound velocity is $c_s^2 = \partial p/\partial e$, we can obtain the sound velocity as
	\begin{equation}
	c_s^2 = \frac{\partial p/\partial T}{\partial e/\partial T} = \frac{4a_{\pm}T^3 - 3cT\phi^2}{12a_{\pm}T^3 - 3cT\phi^2}\,\,.
	\end{equation}
	In the symmetric phase $\phi=0$, then the corresponding sound velocity is
	\begin{equation}
	c_+^2 = \frac{4a_+T^3}{12a_+^2T^3} = \frac{1}{3}\,\,.
	\end{equation}
	However, we have 
	\begin{equation}
	\phi = \sqrt{\frac{-2\lambda\Lambda^2 + 2\Lambda\sqrt{\lambda^2\Lambda^2 - 3\kappa(\mu^2 + cT^2)}}{3\kappa}}\,\,
	\end{equation}
	in the broken phase, and the sound velocity  is
	\begin{equation}
	c_-^2 = \frac{4a_-T^3 - 3cT\phi^2}{12a_-T^3 - 3cT\phi^2}\,\,.
	\end{equation}
	It is obvious that $c_-^2$ can deviate from $1/3$, and $c_-^2$ is also related to the model parameter $\Lambda/\sqrt{\kappa}$ and temperature.
	Our previous study~\cite{Wang:2020nzm} demonstrated the sound velocity as a function of temperature for different cutoff scales in Fig.~2 therein.
	In Table I of Ref.~\cite{Wang:2020nzm}, we found a lower cutoff scale could give a $c_-^2$ that significantly deviates from $1/3$, and $c_-^2$ could even be 0.2317 for $BP_6$.
	Models~\cite{Wang:2019pet, Huang:2017laj} with multiple order-parameter fields could generate two-step phase transition, and 
	the sound velocity of symmetric phase could also deviate from $1/\sqrt{3}$~\cite{Giese:2020rtr,Wang:2020nzm}.  
	For a given phase transition model, when the field-dependent masses of the thermal particles  are comparable with the phase transition temperature, the sound velocities would deviate from $1/\sqrt{3}$ of the pure radiation phase as discussed in Refs.~\cite{Giese:2020rtr,Wang:2020nzm}.

\bibliography{ref.bib}
\end{document}